\documentclass[12pt]{iopart}
\usepackage{graphicx}
\usepackage{dcolumn}
\usepackage{bm}
\usepackage{lipsum}
\usepackage{verbatim} 
\usepackage{chngcntr}
\usepackage{capt-of} 
\usepackage{adjustbox} 
\usepackage{rotating}
\usepackage{braket}
\usepackage{mathtools}
\usepackage[dvipsnames]{xcolor}
\usepackage[hidelinks,colorlinks=true,linkcolor=blue,citecolor=blue,pdfencoding=auto]{hyperref} 
\usepackage{cleveref}
\usepackage[small,labelfont=bf]{caption} 
\usepackage{subcaption}
\begin{document}

\title{Effective field theory analysis of $^3$He-$\alpha$ scattering data}

\author{Maheshwor Poudel}
\address{Institute of Nuclear and Particle Physics, Department of Physics and Astronomy, Ohio University, Athens, OH 45701, USA
}
\ead{mp918716@ohio.edu}

\author{Daniel R. Phillips}%
\address{Institute of Nuclear and Particle Physics, Department of Physics and Astronomy, Ohio University, Athens, OH 45701, USA
}
\ead{phillid1@ohio.edu}
\vspace{10pt}
\begin{indented}
\item[]Jan 2022
\end{indented}

\begin{abstract}
We treat low-energy $^3$He-$\alpha$ elastic scattering  in an effective field theory (EFT) that exploits the separation of scales in this reaction. We compute the amplitude up to next-to-next-to-leading order, developing a hierarchy of the effective-range parameters (ERPs) that contribute at various orders. We use the resulting formalism to analyze data for recent measurements at center-of-mass energies of 0.38--3.12 MeV using the scattering of nuclei in inverse kinematics (SONIK) gas target at TRIUMF as well as older data in this energy regime. We employ a likelihood function that incorporates the theoretical uncertainty due to truncation of the EFT and use Markov Chain Monte Carlo sampling to obtain the resulting posterior probability distribution. We find that the inclusion of a small amount of data on the analysing power $A_y$ is crucial to determine the sign of the p-wave splitting in such an analysis. The combination of $A_y$ and SONIK data constrains all effective-range parameters up to $O(p^4)$ in both s- and p-waves quite well. The asymptotic normalisation coefficients and s-wave scattering length are consistent with a recent EFT analysis of the capture reaction ${}^3$He($\alpha$,$\gamma$)${}^7$Be.
\end{abstract}

%

\section{\label{introduction} Introduction}
The ${}^3$He-$\alpha$ elastic scattering and radiative capture reactions have received a significant amount of recent attention in a number of different theoretical approaches. Cluster models, such as the ones developed in references.~\cite{Tursunmahatov:2012,Tursunov:2020npj}, and optical models (see, e.g., reference~\cite{mohr1993alpha}) describe the low-energy ${}^3$He-$\alpha$ scattering data quite well and reproduce capture data with moderate accuracy. Successful {\it R}-matrix fits of both scattering and capture data have been obtained~\cite{deBoer:2014hha,Thompson:2019uok}. And {\it ab initio} methods now make quantitative statements about the low-energy behavior of the  ${}^3$He($\alpha$,$\gamma$)${}^7$Be reaction~\cite{Neff:2011,DohetEraly:2016}. 

An effective field theory (EFT) suitable for halo or clustered systems has also been applied to these reactions. An EFT is a controlled expansion in a ratio $Q \equiv p_{\text{typ}}/\Lambda$, where $p_{\text{typ}}$ is the low-momentum scale that typifies the scattering 
and $\Lambda$ is the momentum scale at which the theory breaks down---see, e.g., reference~\cite{kaplan1995effective} for an introduction.  In this system the EFT, which we shall refer to here as Halo EFT, is built on the scale separation between the large de Broglie wavelength of the quantum-mechanical scattering or capture process and the small size of the ${}^3$He and $\alpha$ nuclei. In this treatment $^7$Be is a bound state of $^3$He and $\alpha$ nuclei. Such a description is accurate because the $^7$Be nucleus' ground-state binding energy ($E_{\text{GS}}$) of 1.6 MeV \cite{tilley2002energy} and excited state energy ($E_{\text{XS}}$) of 1.2 MeV \cite{tilley2002energy} are small compared to scales associated with the structure of these ${}^7$Be constituents. The proton separation energy $S_\text{p}$ of $^3$He is 5.5 MeV \cite{kinsey1996nudat} and the first excitation energy of the $\alpha$ particle is approximately 20 MeV \cite{kinsey1996nudat}. These energy scales, as well as the sizes of the two helium isotopes, suggest the EFT breaks down for momentum $\Lambda \approx 150$--$200$ MeV. Therefore as long as we use it for ${}^3$He-$\alpha$ scattering at center-of-mass energies ($E_{\rm cm}$) less than about $2.5$ MeV the EFT should have good convergence. This Halo EFT has been applied to the $^3$He($\alpha$,$\gamma$)$^7$Be reaction in references~\cite{higa2018radiative,zhang2020s,premarathna2020bayesian} and was used to fit the scattering phase shifts in references~\cite{higa2018radiative,premarathna2020bayesian}. Readers desiring a more detailed introduction to Halo EFT are referred to the reviews in references~\cite{Hammer:2017tjm,Hammer:2019poc}.

Part of the recent theory interest in these processes is because the capture reaction occurs at center-of-mass energies around 20 keV as part of the solar pp chain that fuels the Sun and Sun-like stars. This reaction has a branching fraction of 16.7\% in the pp chain and leads to subsequent decays of $^7$Be and $^8$B that produce neutrinos~\cite{Adelberger}. This information is used to construct solar models. The ${}^3$He($\alpha$,$\gamma$)$^7$Be reaction rate cannot be directly measured at the energies of relevance for the solar interior. The uncertainties in this reaction rate and in the reaction rate ${}^7$Be($p$,$\gamma$)${}^8$B produce, respectively, uncertainties of 3.9\% and 4.8\% in the flux of ${}^8$B solar neutrinos~\cite{Vinyoles:2016djt}. These nuclear-reaction-rate uncertainties thus limit our ability to precisely test solar models using neutrino data. The ${}^3{\rm He}(\alpha, \gamma){}^7{\rm Be}$ also plays a role in Big Bang nucleosynthesis, although it can be measured directly at the 100--300 keV energies that are relevant in that environment.

Data on the elastic scattering of ${}^3$He and $\alpha$ nuclei does not directly affect input quantities to solar models, but it  definitely helps us understand the capture reaction ${}^3$He($\alpha$,$\gamma$)${}^7$Be phenomenologically. In particular, the experimentally measured cross sections must be extrapolated to lower energies for use in solar models. The energy dependence of that extrapolation is strongly constrained by information from the elastic scattering reaction. 

Halo EFT treatments of elastic ${}^3$He-$\alpha$ scattering data are therefore timely, especially because an elastic scattering experiment was recently performed at center-of-mass energies between 0.38 and 3.13 MeV using the scattering of nuclei in inverse kinematics (SONIK) detector and a ${}^4$He gas target at the TRIUMF facility, Canada. These SONIK data have been published and analysed using the phenomenological {\it R}-matrix theory \cite{laneandthomas} in reference \cite{paneru_thesis}. They join quite a few previous measurements of the elastic scattering ~\cite{miller1958scattering, tombrello1963scattering, barnard1964, spigerandtombrello1967, chuang1971elastic, roos1980he, gorpinich1992elastic, mohr1993alpha}. However, some of these data sets do not quantify their errors well~\cite{mohr1993alpha,miller1958scattering} and others have little low-energy data \cite{chuang1971elastic}. References~\cite{roos1980he} and \cite{gorpinich1992elastic} consider elastic scattering only at higher energies ($E_{\rm lab}>$10 MeV~\footnote{The lab energy is the energy of the projectile nuclei $^3$He in the rest frame of the target $\alpha$. All energies are in this frame in the paper unless otherwise mentioned.}) which is well beyond our region of interest. Similarly, all of the data in reference.~\cite{spigerandtombrello1967} are beyond the highest energy we want to study---and errors are not provided for these data. Reference \cite{tombrello1963scattering} primarily focuses on studying the level structure and resonances in $^7$Be from the elastic scattering and not on differential-cross-section measurements. For these reasons, out of all these older data sets, in the end we just analyse those published in references~\cite{chuang1971elastic} and~\cite{barnard1964}. The latter provides the only substantial data set we fit other than the SONIK data. We call it the ``Barnard data``; it covers a laboratory energy range of 2.5-5.7 MeV.  

We have constructed an EFT to describe ${}^3$He-$\alpha$ elastic scattering data for $E_{\rm lab}=0.5$--$4.5$ MeV. This is the first attempt at describing ${}^3$He-$\alpha$ elastic scattering data (as compared to phase shifts) in EFT. To describe the higher-energy portion of these data at the accuracy we desire the ${\frac{7}{2}}^-$ partial wave must be included in the analysis. But
our EFT does not incorporate the ${\frac{7}{2}}^-$ ${}^7$Be resonance at $E_{\rm lab}=5.22$ MeV~\cite{piluso}. Instead, we employ a phenomenologial treatment of that resonance, based on {\it R}-matrix theory~\cite{laneandthomas}, in order to account for the impact of the ${\frac{7}{2}}^-$ partial wave on observables in the energy range of interest. 

Because the scattering amplitude is computed at a fixed order in the EFT it is accurate up to a given order in the expansion in powers of $Q$. In the Halo EFT used here, an order $\nu$ calculation means that the effective-range function is accurate up to terms of $O(p Q^\nu)$. Assessing how accurate the effective-range function is requires input information about the size of the different effective-range parameters (ERPs). We take ERPs that carry $m$ positive powers of distance, i.e., that are in units ${\rm fm}^m$, to have a size of order $\sim 1/\Lambda^m$, but we find that ERPs with negative distance dimension are generally associated with momentum scales much smaller than $\Lambda$.

This development of an EFT expansion for the ${}^3$He-$\alpha$ scattering amplitude means we can also assign an error of $O(p Q^{\nu+1})$ to the order-$\nu$ calculation. 
A novel feature of our analysis is that we incorporate this theory uncertainty into our likelihood, combining it with the data uncertainties, so that the likelihood accounts for both imperfections in the experimental data and the theoretical model thereof~\cite{wesolowski2019exploring, Wesolowski:2021}. 

We then use Bayesian statistics together with Markov Chain Monte Carlo sampling to construct the posterior probability distribution function (pdf) for the relevant low energy parameters from the SONIK and Barnard data. In order to resolve a bimodality in the resulting pdf we also include a few analysing-power ($A_y$) data in our fit.
The construction of this pdf helps us extract credibility intervals for the ERPs and facilitates propagation of those parameter uncertainties to relevant observables. Bayesian MCMC sampling is also sometimes done for model comparison, i.e.,  to determine the preference of a particular data set for one over another (see, e.g., reference~\cite{premarathna2020bayesian}, where the method is used to compare different EFT power countings) but we have not done such a comparison in this paper.

We have organized the paper in the following way. In section~\ref{EFT_formalism} we will setup $^3$He-$\alpha$ scattering in the EFT framework and introduce the low-energy constants (LECs) of our EFT. In section~\ref{fine_tuning} we will discuss how the LECs are determined by matching the EFT to the effective-range expansion. This results in a mapping from the LECs to the effective range parameters (ERPs). In section~\ref{power_counting}, we propose and check a power counting that organizes contributions to the scattering amplitude into leading-order ($\nu=0$, LO), next-to-leading order ($\nu=1$, NLO), and next-to-next-leading order ($\nu=2$, NNLO) contributions. In section~\ref{error_model} we introduce the model for EFT truncation errors and estimate the parameters $c_{\rm rms}$ and $Q$ that define that error model. We also show that the error model describes the shifts from LO to NLO and NLO to NNLO reasonably well. In section~\ref{data} we give details on the data set used in our Bayesian inference and in section~\ref{sec:fwaves} we describe how we phenomenologically include the effects of $L=3$ partial waves in the analysis. 
In section~\ref{bayesian_formulation} we briefly describe our Bayesian methods and MCMC sampling technique and define the priors on the parameters we are estimating, before presenting our results in section~\ref{results} and giving a summary and outlook in section~\ref{conclusion}.  
\section{\label{EFT_formalism} EFT formalism for ${}^3{\rm He}-\alpha$ scattering}
\noindent In this section we discuss the EFT formalism for ${}^3$He-$\alpha$ scattering. The EFT of this reaction has already been laid out in references~\cite{higa2018radiative,zhang2020s}. The presentation below follows the one in those papers closely, although there are some minor differences that we will point out along the way. 

We denote the spin-$\frac{1}{2}$ field of the $^3$He-nuclei as $\psi$ and the spin-0 field of the $\alpha$ nuclei as $\phi$. 
The invariant amplitude of elastic scattering between $^3$He and $\alpha$ nuclei consists of pure Coulomb scattering at larger distance while at short distance the Coulomb scattering mixes with the scattering from strong interaction. The Lagrangian that describes the Coulomb interaction between the $\psi$ and $\phi$ fields is:
\begin{equation}
\mathcal{L}_C = -\frac{4\pi \alpha_{em} Z_\psi Z_\phi}{|\vec{q}|^2} \int {\rm d}^3y\  e^{-{\rm i}\vec{q}\cdot (\vec{x}-\vec{y})} \phi^\dagger(x) \phi(x) \psi^\dagger(y) \psi(y) \Bigg{|}_{y_0=x_0},    
\end{equation}
where $\alpha_{em}=\frac{1}{137.04}$ is the electromagnetic fine-structure constant, $Z_\psi=Z_\phi=2$ are the number of protons in ${}^3$He and $\alpha$ nuclei respectively and $\vec{q}$ is the momentum transfer. Meanwhile, since the strong force between ${}^3$He and the $\alpha$ particle has a range comparable to the short-distance scale in the problem we treat it as a string of contact interactions. To write the Lagrangian for this force we  adopt the dimer formalism introduced and described in reference~\cite{hammer}. We introduce one dimer field for each of the two channels $J=L\pm\frac{1}{2}$. These will be subscripted as $L^+$ for the $J=L+\frac{1}{2}$ channel dimer and $L^-$ for the other one. Of course, there is only one dimer in the $L=0$ channel, where we do not need the $\pm$ designation because  $J=\frac{1}{2}$ is the only possibility.  
We write the dimer for $L^\pm$ channel as $\mathcal{D}^\pm_L$. The Lagrangian density then is simply the sum of free Lagrangian $\mathcal{L}_\text{free}$, s-wave interaction Lagrangian $\mathcal{L}_S$ that couples the fields to the s-wave dimer, p-wave interaction Lagrangians $\mathcal{L}_{P_{1^+}}$ and $\mathcal{L}_{P_{1^-}}$ and
the Coulomb Lagrangian $\mathcal{L}_C$. 

Up to the order to which we work the free Lagrangian is given by
\begin{multline}
\mathcal{L}_\text{free} = \phi^\dagger \Bigg{(} i\frac{\partial}{\partial t}+\frac{\nabla^2}{2m_\phi} \Bigg{)} \phi 
+ {\psi^\dagger}^a \Bigg{(} i\frac{\partial}{\partial t}+\frac{\nabla^2}{2m_\psi} \Bigg{)}\psi_a \\
 +\sum_{L=0}^1 {({\mathcal{D}^\pm_L})^\dagger}^{\aleph^\pm_L} \Bigg{[} \omega^\pm_L \Bigg{(} i\frac{\partial}{\partial t}+\frac{\nabla^2}{2M} \Bigg{)}
+ \Xi^\pm_L \Bigg{(} i\frac{\partial}{\partial t}+\frac{\nabla^2}{2M} \Bigg{)}^2
+ .... +\Delta^\pm_L \Bigg{]}{(\mathcal{D}^\pm_L)}_{\aleph^\pm_L}
\label{eq:free_lagrangian2}
\end{multline}
We sum over repeated indices in equation~\eqref{eq:free_lagrangian2} and the following equations. The index $a$ runs over $\pm \frac{1}{2}$, the allowed spins of the free $^3$He field. The masses that appear here are $m_\psi=2809.43$ MeV, $m_\phi=3728.42$ MeV, the rest masses of $\psi$ and $\phi$ fields respectively, and $M=m_\psi+m_\phi$. For each channel, $\Delta^\pm_L$ is the bare dimer binding energy, $\omega^\pm_L$ is a dimensionless constant that takes value $\pm 1$ depending upon the sign of effective range, and the constant $\Xi^\pm_L$ has dimensions of inverse energy. These three LECs are adjusted to reproduce the terms of order $1$, $p^2$, and $p^4$ in the effective-range expansion. We will discuss this matching in detail in Sec.~\ref{fine_tuning}. While references~\cite{higa2018radiative,zhang2020s} considered the impact of the $p^4$ term in the effective-range expansion on observables, neither included an operator in the Lagrangian that generated them. Our presentation here goes beyond previous work in this regard. The dots .... represent higher-order terms that carry higher mass dimension than the three terms written explicitly in the dimer piece of ${\cal L}_{\rm free}$.

The index $\aleph^\pm_L$ enumerates the different components of the dimer field. For example, the s-wave dimer with ($J=\frac{1}{2}$, $M_{J_\Theta} \equiv \aleph_0=\pm\frac{1}{2}$) couples to the $\psi$ field with ($j_\psi=\frac{1}{2}$, $m_{j_\psi} = \aleph_0=\pm\frac{1}{2}$) and $\phi$ field with ($j_\phi=0$, $m_{j_\phi}=0$). The respective Lagrangian is given by
\begin{equation}
    \label{s-wave_dimer2}
    \mathcal{L}_S = g_0 \sum_{\aleph_0}{{\mathcal{D}_0}^\dagger}^{\aleph_0} \psi_{{}_{\aleph_0}} \phi + h.c
\end{equation}
where $g_0$ gives the strength of the s-wave interaction.

Regarding p-wave dimers, the piece of the Lagrangian that describes p-wave interactions is:
\begin{equation}
\label{p_dimer}
\mathcal{L}_{P_{1^\pm}} = g^\pm_{1} \sum_{\aleph_1, \lambda, \sigma}{{\mathcal{D}^\pm_1}^\dagger}^{\aleph^\pm_1} 
\braket{\frac{1}{2} \lambda; 1 \sigma| 1^\pm \aleph^\pm_1} \times \Bigg{(}\frac{2m_\psi}{M}\psi_\lambda (i\mathbf{\nabla}_\sigma\phi)
-\frac{2m_\phi}{M}(i\mathbf{\nabla}_\sigma{\psi_\lambda})\phi\Bigg{)} + h.c.
\end{equation}
The interaction strength is given by $g^\pm_1$ and $\langle{\frac{1}{2} \lambda; 1 \sigma}|{1^\pm \aleph^\pm_1}\rangle$ is the Clebsch-Gordan coefficient of the coupling that produces the $P_{1^\pm}$ final state where, for the $1^+(1^-)$ channel we replace $1^+(1^-)$ by $\frac{3}{2}(\frac{1}{2})$. The fields ${\mathcal D}_1^{\pm}$ have components labeled by $\aleph^+_1 = \pm\frac{3}{2}, \pm\frac{1}{2}$ and $\aleph^-_1 = \pm\frac{1}{2}$ respectively. In equation~ \eqref{p_dimer}, the mass terms in front of the derivatives ensure the Galilean invariance of the Lagrangian i.e., they produce the combination of derivatives corresponding to the relative momentum between the two particles.

\begin{figure}
    \centering
    \includegraphics[width=\columnwidth]{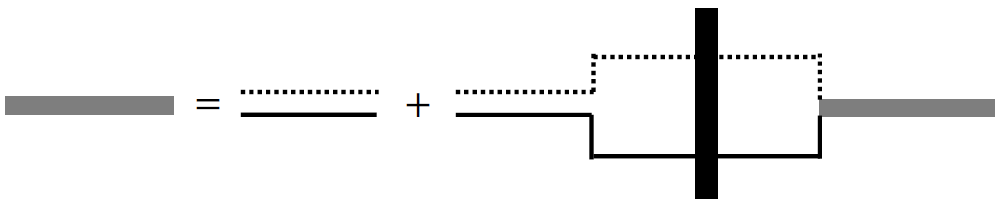}
    \caption[width=\columnwidth]{Dyson equation for the full dimer propagator, represented by the thick gray line. It consists of the bare dimer propagator part (the first term) and the dressing (second term). The bare dimer propagator is represented as the two closed lines; one dashed line representing the $\phi$ field and the other black solid line representing the $\psi$ field. The thick black vertical box in the second figure is the four-point Coulomb amplitude between these fields.}
    \label{full_dimer_image}
\end{figure}

The $\psi$ and $\phi$ fields interact via infinite range Coulomb interaction and the short-range strong interactions induced by the dimers. The combined effect of strong and Coulomb interactions is obtained by computing the dressed dimer propagator. If we denote the dimer self energy by $\Sigma_L$ then the fully dressed scalar dimer propagator for s-wave and p-wave dimers is given by
\begin{equation}
  \label{full_dimer2}
    D^\pm_L(E, \vec{P}) = \Bigg{[}\Delta^\pm_L + \omega^\pm_L \bigg{(}E-\frac{{|\vec{P}|^2}}{2M}\bigg{)}
        +\Xi^\pm_L\bigg{(}E-\frac{{|\vec{P}|^2}}{2M}\bigg{)}^2 + {\rm i}\epsilon-\Sigma^\pm_L(E)\Bigg{]}^{-1}.
\end{equation}
where $\vec{P}$ is the momentum of the dimer, and so is equal to the center-of-mass momentum of the ${}^3$He-$\alpha$ pair that couples to it. The fully dressed dimer propagator is depicted in figure.~\ref{full_dimer_image}. 

\begin{figure}
    \centering
    \includegraphics[width=\columnwidth]{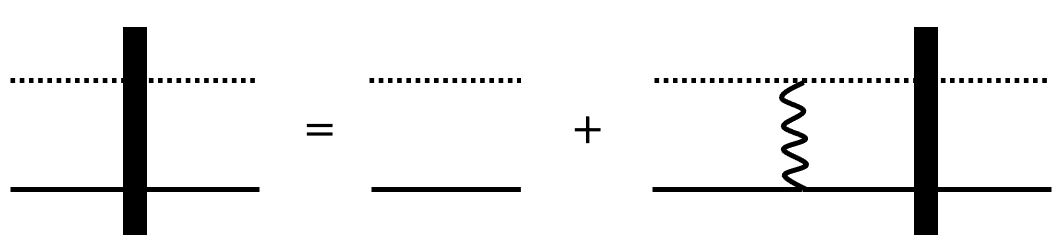}
    \caption{The full Coulomb Green's function: The first term on the right-hand side represents the free Green's function $G_0$. The wavy line is the single photon exchange between the $\psi$ and $\phi$ fields. The black box is the four-point Coulomb amplitude between these fields.}
    \label{coulomb_propagator_image}
\end{figure}

To ease calculating the irreducible self energy in our EFT, we introduce the Coulomb Green's function. We write this propagator in momentum representation as $\tilde{G}_C(E;\vec{p}_2,\vec{p}_1)$:
\begin{equation}
    \label{coulomb_propagator}
    \tilde{G}_C(E;\vec{p}_2,\vec{p}_1) \equiv -G_0(E,{\vec{p}}_2)\mathcal{F}({\vec{p}}_2,{\vec{p}}_1;E)G_0(E,{\vec{p}}_1)
\end{equation}
where $\mathcal{F}$ is the Coulomb four-point function, and $\vec{p}_1$ and $\vec{p}_2$ are, respectively, the incoming and outgoing relative momenta of the ${}^3$He and $\alpha$ particle. The integral equation for this Coulomb Green's function is shown in figure~\ref{coulomb_propagator_image}.
The free propagator $G_0$ is given by 
\begin{equation}
    G_0(E + i \epsilon;\vec{p})= \frac{1} {E-\frac{{|\vec{p}|}^2}{2\mu} + {\rm i}\epsilon},
\end{equation}
where $\mu=m_\psi m_\phi/M $ is the reduced mass of the dimer and we have explicitly pointed out that we will evaluate the free Green's function at $E + i \epsilon$ if $E$ is a real positive number, with $\epsilon \rightarrow 0+$. In co-ordinate representation the Coulomb Green's function $G_C(E; \vec{x}, \vec{x}')$  satisfies the following equation:
\begin{equation}
    -\frac{\nabla_x^2}{2\mu} G_C(E;\vec{x},\vec{x}') + \Biggl(\frac{Z_\psi Z_\phi \alpha_{em}}{x}-E\Biggr)G_C(E;\vec{x},\vec{x}')=\delta^{(3)}(\vec{x}-\vec{x}').
\end{equation}
It has the spectral representation~\cite{kong2000coulomb,Higa:2008dn}
\begin{equation}
    \label{CGF}
    G^{(+)}_C(E; \vec{x}, \vec{x}') = \int \frac{{\rm d}^3q}{(2\pi)^3} \frac{\left(\chi_{\vec{q}}^{(+)}(\vec{x})\right)^* \chi_{\vec{q}}^{(+)}(\vec{x}')}{E-\frac{|\vec{q}|^2}{2\mu} + i\epsilon}
\end{equation}
where $\chi^{(+)}_{\vec{q}} (\vec{x})$ corresponds to outgoing spherical-wave boundary conditions, and so is given by~\cite{kong2000coulomb,Higa:2008dn}:
\begin{equation}
    \label{in-state_sol}
    \chi^{(+)}_{\vec{q}}(\vec{x}) = e^{-\frac{\pi \eta}{2}} \Gamma(1+{\rm i}\eta) \ M(-{\rm i}\eta, 1; {\rm i}qx-{\rm i}\vec{q} \cdot \vec{x}) \ {\rm e}^{{\rm i}\vec{q} \cdot \vec{x}}.
\end{equation}
In equation~\eqref{in-state_sol} $\eta = k_c/p$ with $k_c=Z_\psi Z_\phi \alpha_{em} \mu$, while $M(a, b; z)$ is a confluent hypergeometric function that is also called the Kummer function~\cite{abramowitz1964handbook}. For further details on $\chi^{(\pm)}_{\vec{q}}(\vec{x})$ and $M(a, b; z)$ we refer the reader to references~\cite{kong2000coulomb,Higa:2008dn,abramowitz1964handbook}.

Using equation~\eqref{CGF}, the scalar one-loop self energy for s- and p-wave dimers can be expressed as integrals over the variable $\vec{q}$, since they are proportional to $G_C(E;0,0)$ and derivatives of $G_C$ at the origin. The renormalized self energies, to which we assign a superscript $R$, are then found to be:
\begin{eqnarray}
\label{self_energy_renormalised1}
{\Sigma^{\pm}_L}^R &= \frac{(g^\pm_L)^2}{(2L+1)} \int \frac{d^3p_1 d^3p_2}{(2\pi)^6} \tilde{G}_C(E;\vec{p}_2,\vec{p}_1) (\vec{p}_2 \cdot \vec{p}_1)^L
- {\Sigma^{\pm}_L}^\text{div}\\ 
\label{self_energy_renormalised2}
& = - (2L+1)\frac{\mu (g^\pm_L)^2}{2 \pi} \Bigg{(}\frac{1}{A_L \Gamma(L+1)}\Bigg{)}^2 \Theta_L(\eta; E).   
\end{eqnarray}
In equation~(\ref{self_energy_renormalised2}) $A_L$ and $\Theta_L(\eta;E)$ are given respectively as 
\begin{equation}
    A_L = \frac{\Gamma(2L+2)}{2^L\Gamma(L+1)}
\end{equation}
and
\begin{equation}
    \Theta_L(\eta;E)=2k_cp^{2L}\frac{\Gamma(1+L+{\rm i}\eta)\Gamma(1+L-{\rm i}\eta)}{\Gamma(1+{\rm i}\eta)\Gamma(1-{\rm i}\eta)}H(\eta),
\end{equation}
where $H(\eta)$ is
\begin{equation}
    H(\eta) = \Psi({\rm i}\eta) + \frac{1}{2{\rm i}\eta} - \log {\rm i}\eta,
\end{equation}
where $\Psi({\rm i}\eta)$ is digamma function in $\eta$ defined as in reference~\cite{abramowitz1964handbook}.

Strictly speaking, these condensed forms of the renormalized self energy are only valid for $L\leq 1$. For higher $L$ the particle-dimer vertices  must be constructed out of irreducible tensors of the rotation group of the appropriate rank~\cite{brown2014field,Braun:2018hug,Braun:2018vez}. The self energy for $L \geq 2$ also requires additional renormalization~\cite{Braun:2018hug,Braun:2018vez}. 

The $\Sigma^{\text{div}}$ are the divergent part of the integral in equation~ \eqref{self_energy_renormalised1}. When evaluated using dimensional regularization and the power-law divergence subtraction (PDS) scheme we have, for s- and p-waves separately:
\begin{equation}
    \label{s_wave_divergence}
    \Sigma^{\text{div}}_0 = -\frac{(g_0)^2 \mu \xi}{2\pi}
    + \frac{k_c \mu (g_0)^2}{\pi} \Bigg{[} \frac{1}{\epsilon}-\frac{3}{2}C_E+1+\log\Bigg{(}\frac{\sqrt{\pi}\xi}{2k_c}\Bigg{)} \Bigg{]}
\end{equation}
and
\begin{equation}
    \label{p_wave_divergence}
    {\Sigma^{{\pm}}_1}^\text{div} = {\Sigma^{0\pm}_1}^{\text{div}} + p^2 {\Sigma^{1\pm}_1}^{\text{div}} \\
\end{equation}
with
\begin{equation}
    \label{p_wave_divergence1}
    {\Sigma^{0\pm}_1}^\text{div} = -\frac{\mu (g^\pm_1)^2 k^2_c \xi}{4\pi}\Bigg{(} 1 + \frac{\pi^2}{3} \Bigg{)}
    + \frac{\mu (g^\pm_1)^2 k^3_c}{3\pi} \Bigg{[} \frac{1}{\epsilon} - \frac{3}{2} C_E + \frac{4}{3} + \log \Bigg{(}\frac{\sqrt{\pi} \xi}{2k_c} \Bigg{)}\Bigg{]}
\end{equation}
and
\begin{equation}
    \label{p_wave_divergence2}
    {\Sigma^{1\pm}_1}^\text{div} = -\frac{\mu (g^\pm_1)^2 \xi}{4\pi}
    + \frac{\mu (g^\pm_1)^2 k_c}{3\pi}\Bigg{[} \frac{1}{\epsilon} - \frac{3}{2} C_E + \frac{4}{3} + \log \Bigg{(}\frac{\sqrt{\pi} \xi}{2k_c} \Bigg{)}\Bigg{]} 
\end{equation}
Note that the linear divergence in the p-wave self energy is proportional to the energy. In equations~\eqref{s_wave_divergence}, \eqref{p_wave_divergence1} and \eqref{p_wave_divergence2}, $\xi$ is the scale introduced by PDS~\footnote{In this work we use $\mu$ to denote the ${}^3$He-$\alpha$ reduced mass.}, which also preserves the dimensionality of the integrals when generalised to $D=4-\epsilon$ dimensions where $\epsilon$ is a very small number. The quantity $C_E$ is the Euler-Mascheroni constant. 

\begin{figure}
    \centering
    \includegraphics[width=\columnwidth]{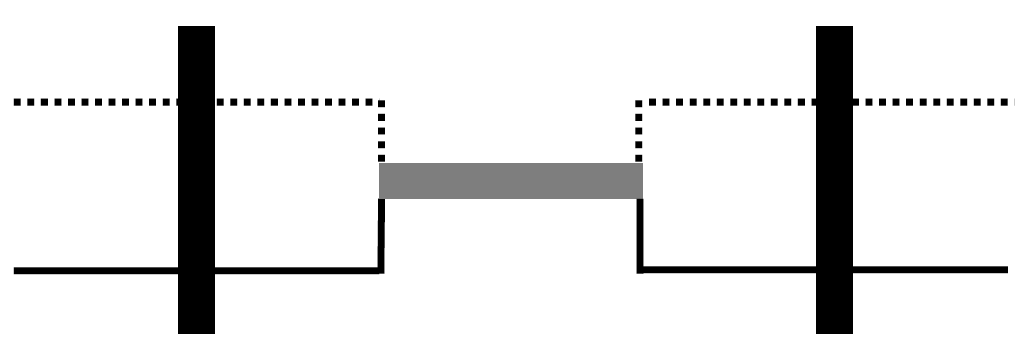}
    \caption{Invariant amplitude, $t$, given by equation~\ref{t_matrix} often denoted $t^{\rm CS}$ in other literature.}
    \label{t_matrix_image}
\end{figure}

It is then straightforward to calculate the piece of the amplitude to which strong forces contribute (i.e. the piece of $t$ other than the pure Coulomb part). It is found by dressing the vertices produced by equations ~\eqref{s-wave_dimer2} and \eqref{p_dimer} with Coulomb Green's function and attaching those vertices to the full dimer propagator, see
figure~\ref{t_matrix_image}. We call the resulting Coulomb-strong interference amplitude, often denoted $t^{CS}$ in the literature, ``the invariant amplitude" henceforth.
The Feynman graph in figure~\ref{t_matrix_image} evaluates to
\begin{equation}
    \label{t_matrix}
    t^\pm_L(E) = (2L+1)^2 \ (g^\pm_L)^2 \text{e}^{2i\sigma_L} C^2_L(\eta) D^\pm_L(E) p^{2L} P_L(\hat{p} \cdot \hat{p}')
\end{equation}
In equation~\eqref{t_matrix} $\vec{p}(\vec{p}')$ is the relative momentum of the incoming(outgoing) particles in the center of mass frame s.t $|\vec{p}| = |\vec{p}'| = p$, $P_L$ is the Legendre polynomial, the quantity $\sigma_L$ is the Coulomb phase shift which is given by
\begin{equation}
    \label{coulomb_phase_shift}
    \text{e}^{2 {\rm i}\sigma_L}= \frac{\Gamma(1+L+{\rm i}\eta)}{\Gamma(1+L-{\rm i}\eta)}
\end{equation}
and the quantity $C_L(\eta)$ is given by
\begin{equation}
    \label{CL}
    C_L(\eta) = \frac{2^L}{\Gamma(2L+2)}\exp{(-\pi \eta/2)}\ |\Gamma(1+L+{\rm i}\eta)|.
\end{equation}
The parameters of the invariant amplitude are the LECs $\Delta_L$, $g_L$, and $\Xi_L$. The task is now to relate these parameters to quantities that can be extracted from experiment. 
\section{Determination of low-energy constants by matching to the effective-\\range expansion}
\label{fine_tuning} 
The alert reader will already have observed that equation~\eqref{t_matrix} has the same form as is obtained when the Coulomb-modified effective-range expansion (CM-ERE) given by equation~\eqref{CMERE} is plugged in scattering {\it T}-matrix given by equation~\eqref{t_matrix_PWA}. We followed the procedures outlined in reference~\cite{kong2000coulomb} for our case. The EFT contains only minimal assumptions about the ${}^3$He-$\alpha$ dynamics: rotational invariance, unitarity, analyticity of the amplitude, and the presence of a short-range strong interaction. Since the effective-range expansion is based on the same set of assumptions, it is not surprising that the EFT reproduces it. The EFT Lagrangian is expressed as an expansion in powers of $p^2$, so, at a given order in the EFT, the ERE is reproduced up to the corresponding order of $p^2$. In this section we perform the matching between the EFT amplitudes obtained in the previous section and the amplitudes of effective-range theory, thereby deriving relationships between the LECs of the EFT and the parameters of the ERE. 

In effective-range theory the amplitude  associated with ${}^3$He-$\alpha$ scattering in the $(L,J=L\pm \frac{1}{2})^\text{th}$ channel, $T_L^{\pm}$, takes the form~\cite{bethe1949theory,hamilton1973coulomb}:
\begin{equation}
\label{t_matrix_PWA}
    T^\pm_L(E + {\rm i} \epsilon)= -(2L+1)\frac{2\pi}{\mu}
    \frac{A^2_L C_L^2 \text{e}^{2i\sigma_L} p^{2L} P_L(\hat{p} \cdot \hat{p}')}{A^2_L C_L^2 p^{2L+1} (\cot \delta^\pm_L - {\rm i})},
\end{equation}
where the quantities $\delta^\pm_L$ are the phase shifts for the channels $L^\pm$. The phase shift for the $\pm$ channels in the $L$th partial wave are, in turn, given by 
\begin{equation}
\label{CMERE}
A^2_L C^2_L p^{2L+1}(\cot \delta^\pm_L - {\rm i})  = 2k^{2L+1}_c K^\pm_L(E)
    -\frac{\Theta_L}{(\Gamma(L+1))^2},
\end{equation}    
where
\begin{equation}
    \label{K-function}
    K^\pm_L = \frac{1}{2k_c^{2L+1}}\Bigg{(}-\frac{1}{a^\pm_L} + \frac{1}{2} r^\pm_L p^2 + \frac{1}{4} P^\pm_L p^4 + O((p^2)^3) \Bigg{)}
\end{equation}
is the effective-range function. Equations~\eqref{CMERE} and \eqref{K-function} relate the phase shifts to the coefficients of powers of $p^2$ in the expansion of the function $K$. $K(E)$ is analytic in $p^2$ for $|\vec{p}|< 1/R$, where $R$ is the range of the strong interaction.  The coefficients of $K$'s Taylor series in $p^2$ are (apart from numerical factors) the effective range parameters (ERPs). To obtain phase shifts from ERPs, the polynomial $K$-function is truncated at a suitable order. We will investigate this truncation further in section~\ref{power_counting}.

With the phase shifts in hand the differential cross section---which is the main physical observable in $^3$He-$\alpha$ scattering---can be obtained. This is done using the formula~\cite{{critchfieldanddodder, spigerandtombrello1967}}:
\begin{equation}
    \label{CS_formula}
    \frac{d\sigma}{d\Omega}=|f_c(\theta)|^2+|f_i(\theta)|^2,
\end{equation}
where
\begin{multline}
    \label{fc}
    f_c(\theta)= -\frac{\eta}{2p}\csc^2(\theta/2) \exp[{\rm i}\eta \log(\csc^2(\theta/2))] \\ + \frac{1}{p} \sum_{L=0}^{\infty} \exp(2{\rm i}(\sigma_L-\sigma_0)) P_L(\cos \theta) \Bigg{[} \frac{L+1} {\cot{\delta^+_L}-{\rm i}} + \frac{L}{\cot{\delta^-_L-{\rm i}}}\Bigg{]},
\end{multline}
and
\begin{equation}
    \label{f_i}
    f_i(\theta)= \frac{1}{p} \sum_{L=1}^{\infty} \exp(2{\rm i}(\sigma_L-\sigma_0)) \sin\theta \frac{{\rm d}P_L(\cos \theta)}{{\rm d}\cos\theta} \Bigg{[} \frac{1}{\cot{\delta^-_L-{\rm i}}}
   - \frac{1}{\cot{\delta^+_L-{\rm i}}} \Bigg{]}  .
\end{equation}
In equations~\eqref{CS_formula}-\eqref{f_i}, $\theta$ is the scattering angle in the centre-of-mass frame and $P_L$  is the $L$th Legendre polynomial. The difference of Coulomb phase shifts that appears here is computed using the formula~\cite{abramowitz1964handbook, barata2009coulomb}:
\begin{equation}
    \sigma_L-\sigma_0=\sum_{m=1}^L \arctan\left(\frac{\eta}{m}\right).
\end{equation}

In addition to the differential cross section we also consider the analysing power $A_y(\theta)$. Measurement of this observable requires the ability to polarize the ${}^3$He nuclei perpendicular to the scattering plane, i.e., along the $\hat{y}$ axis. After measuring the differential cross section for two polarisations in the $+\hat{y}$ and $-\hat{y}$ directions we then construct:
\begin{equation}
    A_y(\theta) \equiv \frac{\frac{d \sigma}{d \Omega}_{\uparrow} -\frac{d \sigma}{d \Omega}_{\downarrow}}
    {\frac{d \sigma}{d \Omega}_{\uparrow} + \frac{d \sigma}{d \Omega}_{\downarrow}}.
\end{equation}
In terms of the amplitudes $f_c$ and $f_i$, $A_y$ is given by~\cite{goldberger2004collision}:
\begin{equation}
    \label{Ay}
    A_y(\theta)= \frac{-2 \Im(f_c f^*_i)}{|f_c|^2+|f_i|^2}.
\end{equation}

We now match the CM-ERE amplitudes $T_L^{\pm}$ to the EFT amplitudes $t_L^{\pm}$ for the cases $L=0$ and $1$, i.e., demand
\begin{equation}
    \label{fine_tune}
    t_0 = T_0;  \quad   t^\pm_1 = T^\pm_1.
\end{equation}
This is done by choosing the LECs such that particular ERPs are obtained. 
The values of the LECs that accomplish this in the s-wave are
\begin{eqnarray}
    \label{fine_tune_a0}
    -\frac{1}{a_0} &=& -\frac{2\pi}{\mu (g_0)^2}( \Delta_0 - \Sigma^{\text{div}}_0)\\
    \label{fine_tune_r0}
    r_0 &=& -\frac{2\pi \omega_0}{\mu^2 (g_0)^2},\\
    P_0 &=& -\frac{2\pi \Xi_0}{\mu^3 (g_0)^2}.
\end{eqnarray}
In the p-wave the LECs responsible for ensuring $t_1^{\pm}=T_1^{\pm}$ are
\begin{eqnarray}
    \label{fine_tune_a1}
    -\frac{1}{a_1^\pm} &=&  -\frac{6\pi}{\mu (g^\pm_1)^2} (\Delta^\pm_1 - {\Sigma^0_1}^{\text{div}}) \\ \label{fine_tune_r1}
    r^\pm_1 &=& -\frac{6\pi}{\mu^2(g^\pm_1)^2} ( \omega^\pm_1 - 2\mu {\Sigma^1_1}^{\text{div}}),\\
    P_{1}^{\pm} &=& -\frac{6\pi \Xi^\pm_{1}}{\mu^3(g^\pm_1)^2}.
\end{eqnarray}
The resulting amplitudes $T^\pm_1$ have poles at momentum $p={\rm i}\gamma^\pm_{1}$ (equivalently $\eta^\pm_{B}=k_c/\gamma^\pm_{1}$) corresponding to the bound state energy $B^\pm=-E^\pm=({\gamma^\pm_{1}})^2/2\mu$. Here $\gamma^\pm_{1}$ are the binding momenta corresponding of the shallow p-wave bound states in, respectively, the $\frac{3}{2}^-$ and $\frac{1}{2}^-$ channels.
The $\frac{3}{2}^-$ is the ground state and lies 1.5866 MeV~\cite{tilley2002energy} below the ${}^3$He-$\alpha$ threshold, while the first excited state $\frac{1}{2}^-$ is 0.43 MeV~\cite{tilley2002energy} above the ground state and is bound with respect to the two-particle threshold by 1.1575 MeV~\cite{tilley2002energy}. Setting the denominator of equation~\eqref{t_matrix_PWA} to zero for the corresponding momenta so that there are  poles at the empirical bound state energies gives:
\begin{equation}
    \label{constraint2}
    \frac{1}{a^\pm_{1}}= -\frac{1}{2} r^\pm_{1} ({\gamma^\pm_{1}})^2+  P^\pm_{1} ({\gamma^\pm_{1}})^4 + 2 k_c (({\gamma^\pm_{1}})^2 - k^2_c) H(-{\rm i} \eta^\pm_{B}).
\end{equation}
In addition to the physical poles that are the solutions of equation~\eqref{constraint2} the p-wave amplitude can have unphysical poles. These are solutions of equation~\eqref{constraint2} that occur at values of $\gamma_1^{\pm}$ which are large enough that they are outside the radius of convergence of the EFT/effective-range expansion. Consequently, such poles cannot be regarded as real consequences of the the theory. Because they only occur at energies outside the energy range where we apply our EFT, they do not produce structures in the cross section in the region where we examine data.  Unphysical poles that do not affect the EFT's predictions can also occur in the s-wave amplitude.

We also have a useful relationship between two other constants associated with each bound state: the asymptotic normalisation coefficient (ANC) and the wavefunction renormalisation coefficient (WRC). The product of the ANC, $\mathcal{C}^\pm_L$, and the Whittaker function, $W_{-\eta_B,L+\frac{1}{2}}(2\gamma r)$, gives the asymptotic form of the bound-state solutions, $w^\pm_L(r)$, to the radial Schr\"odinger equation:
\begin{equation}
    \label{ANC_eq}
    w^\pm_L(r) \stackrel{r \rightarrow \infty}{\longrightarrow} \mathcal{C}^{\pm}_L W_{-\eta_B,L+\frac{1}{2}}(2\gamma^\pm r).
\end{equation}
Meanwhile, the WRC $\mathcal{Z}^\pm_L$ is the residue of the renormalized dimer propagator which is given by
\begin{equation}
\label{WRC}
\frac{1}{\mathcal{Z}^\pm_L} = \frac{\partial (1/D^\pm_L(E))}{\partial E} \Bigg{|}_{E=-B^{\pm}}. 
\end{equation}
For p-waves we have the relationship between the ANC and the WRC as~\cite{higa2018radiative,ZNP2013}:
\begin{equation}
    \label{p-wave_ANC}
    (\mathcal{C}^\pm_{1})^2 = \frac{(g_1^{\pm})^2}{3\pi} ({\gamma^\pm_{1}})^2 \mu^2 (\Gamma(2+\eta^\pm_{B}))^2 \mathcal{Z}^\pm_{1}.
\end{equation}
But, because of the matching and equation~(\ref{t_matrix}), the derivative in equation~\eqref{WRC} can be expressed as a derivative of the CM-ERE amplitude. So that derivative, evaluated at the bound-state pole is, in turn, related to the ANC. This allows us to write the p-wave effective ranges $r^\pm_{1}$ in terms of the corresponding ANCs:
\begin{equation}
\label{constraint1}
    r^\pm_{1} = -\frac{2 ({\gamma^\pm_{1}})^2 (\Gamma(2+\eta^\pm_{B}))^2}{{{\mathcal C}_{1}^{\pm}}^2}
    + P^\pm_{1} ({\gamma^\pm_{1}})^2 + 4 k_c H(-{\rm i}\eta^\pm_{B})
    + {\rm i} 2 k_c \eta^\pm_{B} \Bigg{(}1- {(\eta^\pm_{B})}^2\Bigg{)} \frac{{\rm d}H(\eta)}{{\rm d}\eta} \Bigg{|}_{\eta=-{\rm i} 
\eta^\pm_{B}}.
\end{equation}
It is preferable to express the $r_{1^\pm}$ in terms of ANCs rather than WRCs, because ANCs are directly related to the bound-state solution of the Schr\"{o}dinger equation.

\section{\label{power_counting} Power counting}
Power counting is an indispensable part of EFT. The EFT expansion of an observable $y$ in a small parameter $Q$ is ~\cite{wesolowski2019exploring, weinberg1979phenomenological, epelbaum2009modern}
\begin{equation}
    \label{power_counting_eqn}
    y(p,\theta) = y_{\text{ref}}(p,\theta) \sum_\nu c_\nu(p,\theta) \ Q^\nu.
\end{equation}
The counting index $\nu$ in general depends on the fields in the effective theory, the number of derivatives, and the number of loops. The index $\nu$ is bounded from below and the contribution corresponding to this lowest value is called LO. The immediate correction corresponding to the next largest value of $\nu$ is called NLO and so on and so forth.

The expansion parameter $Q$ is one key ingredient in equation~ \eqref{power_counting_eqn}. It is the ratio of the typical momentum ($p_{\rm typ}$) of the collision to the breakdown scale ($\Lambda$) of the EFT i.e. $Q = p_{\text{typ}}/\Lambda$. We consider the typical momentum of the collision to be $p_{\rm typ}=\max\{q,p\}$, where  $q=2p\sin(\theta/2)$ is the momentum transfer of the scattering reaction. The bulk of SONIK data is above the lab momentum 60 MeV. At a lab momentum of approximately 90 MeV there is a resonance effect of $\frac{7}{2}^-$ channel. We include this effect but do so only phenomenologically (see section~\ref{F-wave_phase}). Thus in our work we have $p$ between 60 MeV to 90 MeV, so $p$ is above the Coulomb momentum scale $k_c=47$ MeV. We have built an EFT that is valid in this region and  call it the `validity region' henceforth.  The first excitation momentum of $\alpha$ nuclei is $\approx200$ MeV and is taken as the hard momentum/breakdown scale. This is also the scale set by the range of the ${}^3$He-$\alpha$ interaction~\cite{zhang2020s}. It may be surprising that the momentum scale associated with the breakup of ${}^3$He into a proton and a deuteron does not set $\Lambda$. We will show in the next section that the choice $\Lambda=200$ MeV produces a regular EFT convergence pattern for observables, and so, empirically, this is the correct breakdown scale for the EFT description of ${}^3$He-$\alpha$ scattering. Our EFT therefore has an expansion parameter ranging between $\approx$ 0.30-0.45 in the validity region.

Our power counting scheme is based on the observed sizes of ERPs fit to data, viz: $\frac{1}{a_0}, r_0, \frac{1}{a_1^{\pm}}, r_1^{\pm}, P_1^{\pm}$, and the Coulomb characteristic pieces of the denominator of $T$: $\zeta_0 = 2k_c \mathcal{R}(H(\eta))$ and $\zeta_1 = 2k_c p^2 (1+\eta^2) \mathcal{R}(H(\eta))$. To facilitate understanding these scales we already provide the maximum {\it a posteriori} (MAP) values at NNLO for each parameter in table~\ref{N2LO_trunc_paneru+ay}.

\begin{figure*}
    \centering
    \includegraphics[width=\textwidth]{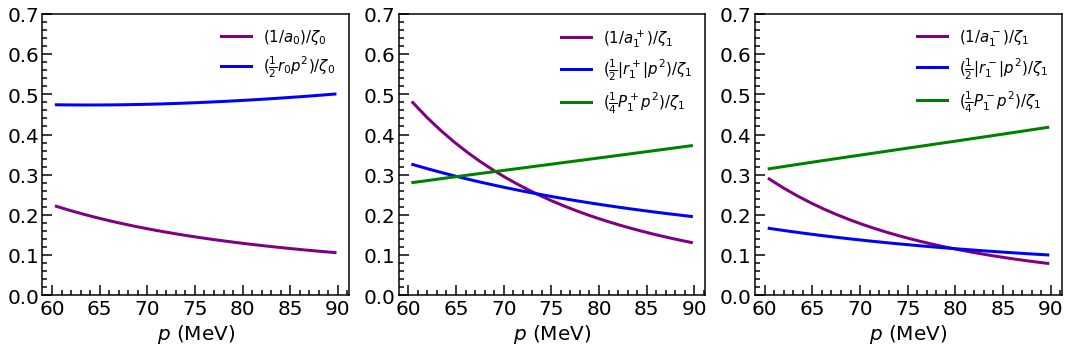}
    \caption{Comparison of scales of various terms in CM-ERE to determine the hierarchy of power counting: The term $\zeta_0 = 2k_c\mathcal{R}(H(\eta))$ and $\zeta_1 = 2 k_c p^2 (1+\eta^2)\mathcal{R}(H(\eta))$.  The left panel is for s-waves, the middle panel for the $\frac{3}{2}^-$ channel, and the right panel for the $\frac{1}{2}^-$ channel.}
    \label{power_counting_figure}
\end{figure*}

In the validity region the Sommerfeld parameter $\eta$ ranges between about $0.5$ and $1$. At these values the pure-Coulomb amplitude and CM-ERE amplitude are both important; neither can be treated perturbatively. In figure~\ref{power_counting_figure} we consider the different contributions to the denominator of the CM-ERE amplitude. The comparison of the Coulomb characteristic quantity $\zeta_0$ and $\zeta_1$ with other terms in the CM-ERE (see figure~\ref{power_counting_figure}) suggests they are markedly bigger than all the terms in $K$ that are proportional to ERPs.  Hence at LO we introduce only the contributions proportional to the $H$-function for both s-and p-wave channels. We say that $\zeta_0$ is of order $\{k_C,p\}$ and $\zeta_1 \sim p^3$. This is the analog of the unitarity limit in the neutral-particle case since all ERPs disappear from the amplitude. Although, in this case, the limit does not yield a scale-invariant amplitude, since the scale $k_C$ persists. 
\begin{table}[b]
    \caption{Hierarchy of power counting in our EFT.}
    \label{power_counting_table}
    \centering
    \begin{tabular}{cccc}
    \br
    Order & s-wave & p-wave & $\nu$ \cr
    \mr
    LO & $-$ & $-$ & 0\cr
    NLO & $r_0$ & $r^+_1$, $P^\pm_1$ & 1\cr
    NNLO & $\frac{1}{a_0}$ & $\frac{1}{a^\pm_1}, r^-_1$ & 2 \cr
    \br 
    \end{tabular}
\end{table}
The terms in the effective-range expansion are corrections to this limit. But to organize them we need to discuss the sizes of the various ERPs.
We categorize the ERPs into two categories: momentum-scale ERPs (that carry positive powers of momentum) and length-scale ERPs (that carry positive powers of length). Table~\ref{N2LO_trunc_paneru+ay} shows that all the momentum-scale ERPs ($\frac{1}{a_0}, \frac{1}{a^\pm_1}, r^\pm_1$) are unnaturally small compared to naive dimensional analysis based on powers of $\Lambda$, while length-scale ERPs ($r_0, P^\pm_1$) tend to be natural when expressed in powers of $1/\Lambda$. We now discuss the scales of these parameters separately for s-and p-waves.
\begin{table*}
\caption{\label{N2LO_trunc_paneru+ay}NNLO extraction of parameters for a combination of SONIK dataset ({\it S}) and Boykin $A_y$ dataset truncated at 4.34 MeV. This sampling uses 407 data points to estimate 16 parameters. The MAP values of the parameters yield a $\chi^2/N_{\rm d.o.f.}$ of 0.86.}
\lineup
\centering
\begin{adjustbox}{width=1\textwidth}
\small
\begin{tabular}{@{}*{10}{l}}
\br
Data & $a_0$ (fm) & $r_0$ (fm) & $a^+_1$ (fm$^3$) & $r^+_1$ (fm$^{-1}$) & $P^+_1$ (fm) & $a^-_1$ (fm$^3$) & $r^-_1$ (fm$^{-1}$) & $P^-_1$ (fm) & $\Gamma_{{\frac{7}{2}}^-}$ (keV) \cr
\mr
$SA_y$ & 60(6) & 0.78(2) & 172(5) & \-0.082(4) & 1.50(4) & 288(15) & \-0.044(6) & 1.65(6) & 159(7) \cr
\br
\end{tabular}
\end{adjustbox}
\end{table*}

Regarding s-waves, the term $1/a_0$ is a very small momentum. We take it as $Q^3 \Lambda$. The s-wave effective range $r_0$ scales naturally as $1/\Lambda$. Since $p_{\rm typ}/\Lambda \equiv Q$, for s-waves, we consider the contribution from $r_0$ at NLO and from $1/a_0$ at NNLO. 

Regarding p-waves, we will rely on the relative importance of each term as suggested from the second and third panels of figure~\ref{power_counting_figure}. We consider both p-wave shape-parameter terms, $\frac{1}{4} P^\pm_1 p^4$, at the same order, even though $\frac{1}{4} P^+_1 p^4$ is a bit smaller than its $J=1/2$  counterpart. In the validity region all other terms in $K_1(E)$ are significantly suppressed which corroborates taking only $\frac{1}{4} P^\pm_1 p^4$ at NLO. We note that all other p-wave terms behave opposite in sign to these NLO effects. For both channels, the terms involving $1/a_1$ and $r_1$ change their relative importance, intersecting in the vicinity of 75 MeV. However, the negative contribution of the term $\frac{1}{2}r_1^+ p^2$as compared to $1/a_1^+$ is steadier and on average larger. Thus in the $J=\frac{3}{2}^-$ channel, we take the contribution from the term $\frac{1}{2}r_1^+ p^2$ at NLO and consider that of $1/a_1^+$ only at NNLO. Both $1/a_1^-$ and $\frac{1}{2}r_1^- p^2$ are considered to be NNLO contributions. Note that although $1/a^\pm_1$ appear larger-than-NNLO at momenta of order 60 MeV, in this region the cross section is still dominated by the Coulomb amplitude. The hierarchy of terms in the CM-ERE matters most at the upper end of the validity region. 

We have summarized this hierarchy of parameters in table ~\ref{power_counting_table}. There are other hierarchies which could be deemed suitable. For example, we demoted $r^+_1$ from NLO to NNLO and got very similar overall results.

In obtaining the cross sections in figure~\ref{fig:validation} we have included f-waves in each result. Removing the f-waves only alters the cross section beyond $120^\circ$ in the last panel corresponding to 
 $E_{\text{lab}} = 4.3$ MeV. 

We also emphasize that relations equations~\eqref{constraint2} and ~\eqref{constraint1} are not obeyed at LO and NLO, i.e., bound-state properties are not reproduced at those orders in this organization of the problem. Thus, while the NNLO amplitude used here is equivalent to that employed at NLO in reference~\cite{zhang2020s}, the organization at lower orders differs from the one employed for capture in that work. 

\section{Error model}
\label{error_model}
We now follow Refs.~\cite{wesolowski2019exploring,Wesolowski:2021,Furnstahl:2015} and use the expansion of equation~\eqref{power_counting_eqn} to build an error model that accounts for the impact that N$^3$LO terms can be expected to have on the cross section. Organizing the differential cross section according to equation~\eqref{power_counting_eqn} implies that the first omitted term in the series, whose order we denote by $\nu=\nu_{\rm max} + 1$, will induce an error
\begin{equation}
    \Delta y(p,\theta)=y_{\rm ref}(p,\theta) c^{\rm rms} Q^{\nu_{\rm max} + 1}.
    \label{eq:errormodel}
\end{equation}
Our main results are EFT differential-cross-section calculations which are at NNLO where $\nu_{\rm max}=2$. But we also use equation~\eqref{eq:errormodel} to estimate the error at LO ($\nu_{\rm max}=0$) and NLO ($\nu_{\rm max}=1$). 

In equation~\eqref{eq:errormodel} $c^{\rm rms}$ is the root-mean-square (rms) value of the coefficients $c_1$, \ldots, $c_{\nu_{\rm max}}$.
We reconstruct $c_\nu$ from predictions of $y$ at orders $\nu_{\rm max}=0,1,2$. When we do this we expect to find 
expansion coefficients $c_\nu$ in equation~\eqref{power_counting_eqn} that are of order unity and roughly similar in size across the different orders $\nu$.
 
For our work, we have chosen the cross section at LO as $y_{\text{ref}}$. This renders $c_0$=1 trivially, so we do not use $c_0$ any further in our analysis.   For the  kinematic point $(p_\gamma,\theta_\eta)$ we then have
\begin{equation}
    \label{c_nu}
    {c_\nu(p_\gamma,\theta_\eta)} = \frac{(y(p_\gamma, \theta_\eta))_\nu-(y(p_\gamma, \theta_\eta))_{\nu-1}}{(y_{\text{ref}}(p_\gamma,\theta_\eta)) Q_{\gamma \eta}^{\nu}}.
\end{equation}

\begin{figure}[t]
    \centering
    \includegraphics[scale=0.5]{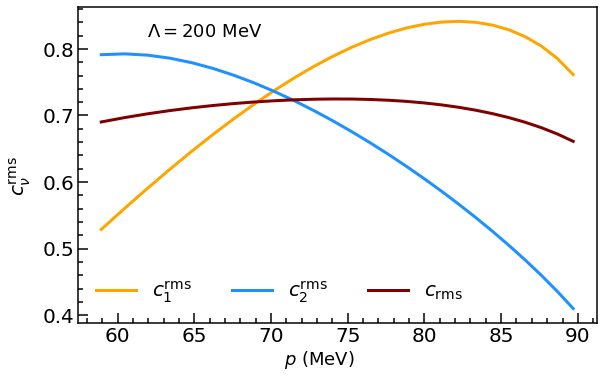}
    \caption{The rms coefficients for the EFT expansion of the ${}^3$He-$\alpha$ cross section. Note that these coefficients are generated using the MAP ERP values obtained in the NNLO fit and given in table~\ref{N2LO_trunc_paneru+ay}. $c_1^{\rm rms}$ is then the NLO-to-LO shift (appropriately rescaled) at those parameter values and $c_2^{\rm rms}$ is the corresponding NNLO shift. Our EFT is valid in the momentum region between 60-90 MeV, so that is the region shown here. Note the suppressed zero scale.}
    \label{crms}
\end{figure}

The rms coefficients in the validity region are shown in figure~\ref{crms}. Here the root-mean-square is computed over angles: the $c_\nu^{\text{rms}}(p)$s shown in figure~\ref{crms} are:
\begin{equation}
    c_\nu^{\text{rms}}(p_\gamma) = \sqrt{ \frac{1}{N_\eta} \sum_\eta c^2_\nu(p_\gamma,\theta_\eta)},
\end{equation}
where $N_\eta$ is the total number of scattering angle points. 
Figure~\ref{crms} shows that organizing the importance of the ERPs according to the hierarchy in table \ref{power_counting_table} produces differential-cross-section EFT coefficients $c_\nu$ that are of order unity. 

A more sophisticated version of this analysis would involve fitting Gaussian Processes to describe the coefficients $c_1(p,\theta)$ and $c_2(p,\theta)$~\cite{Melendez:2019}. Such an analysis would have the advantage that we learn about the way that EFT errors are correlated in momentum and angle. However, since this is a simple EFT with only two low-energy scales, here we expect $c_\nu^{\text{rms}}(p)$ to be fairly constant throughout the validity region, and this is indeed what we observe in figure~\ref{crms} (note the suppressed-zero scale in the figure). This justifies the error model \eqref{eq:errormodel}, with the error assigned there correlated across different kinematic points (cf. equation~(\ref{theory_cov}) below). 

Figure~\ref{crms} also shows that the choice $\Lambda=200$ MeV results in $c_1$ and $c_2$ being of similar size. Lower choices of $\Lambda$ produce coefficients that get smaller as $\nu$ increases, while higher ones result in coefficients that grow with $\nu$~\cite{Melendez:2017}.

\begin{figure*}[t!]
    \centering
    \includegraphics[width=\textwidth]{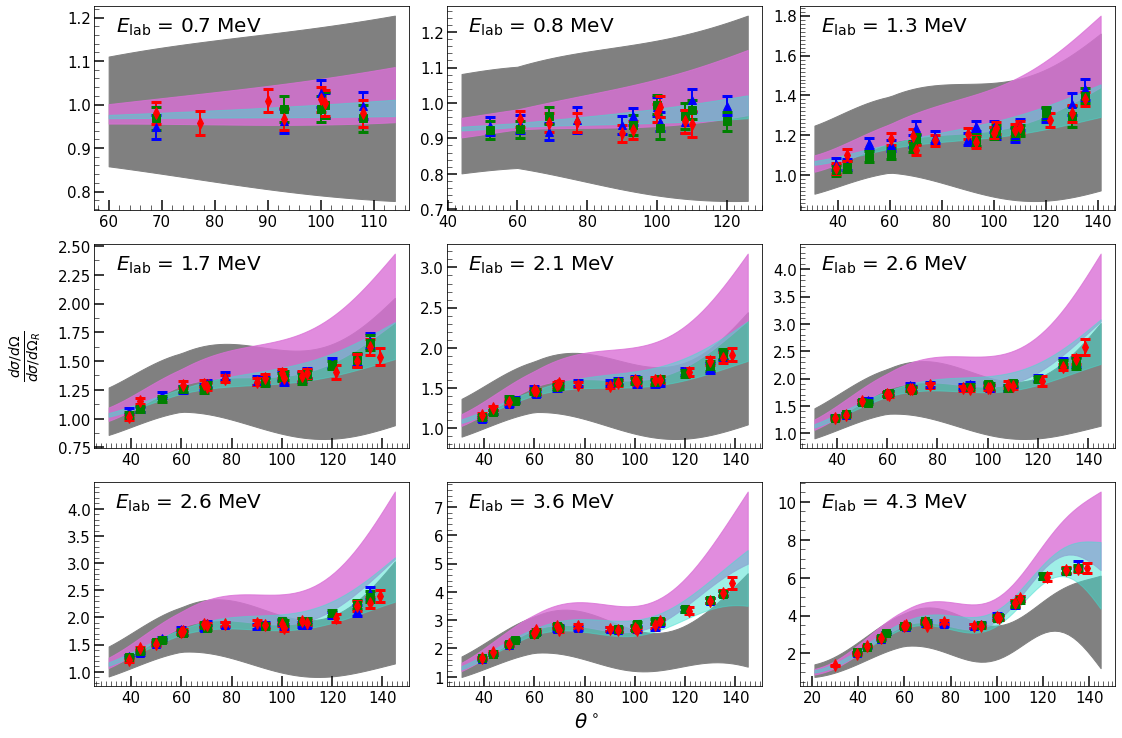}
    \caption{Bands representing the Halo EFT result for and truncation uncertainty in the differential cross section (divided by the Rutherford cross section) at different orders: LO (gray), NLO (pink) and NNLO (cyan) for nine different energies. The SONIK data is also shown in each energy bin. Blue triangles are from IR1, green squares from IR2, and red  diamonds from IR3. At a given beam energy there is a slight difference in the scattering energy for the three different interaction regions, but they are binned together here.}
    \label{fig:validation}
\end{figure*}

Taking the rms of $c_\nu^{\text{rms}}(p)$ over $\nu$ and over the validity region yields  $c^{\rm{rms}}=0.7$ for our analysis. In figure~\ref{fig:validation} we plot the LO, NLO, and NNLO EFT predictions and the associated error of equation~\eqref{eq:errormodel} at each order for the different energy bins measured in the SONIK experiment. We also include the SONIK data in the plot. The LO error bars constructed according to equation~(\ref{eq:errormodel}) typically encompass the NLO result, and the NLO error bars cover the NNLO curves too. Since the NNLO result is fitted to these data it is not entirely surprising that it is consistent with the SONIK data within the EFT uncertainty, but the good agreement within the (small) NNLO uncertainty is reassuring. This validates the hierarchy of our power counting and the error model we have adopted. We point out explicitly that the assignment $Q=\max\{q,p\}/\Lambda$ 
is key to these good results. The EFT uncertainty becomes larger as the momentum transfer increases, i.e., at backward angles, as is most easily seen in the $4.3$ MeV panel in figure~\ref{fig:validation}. In particular, in that panel the NLO band and NNLO band become of comparable size at backward angles. This justifies the truncation of SONIK data at $E_{\rm lab} = 4.3$ MeV and the choice of $\Lambda = 200$ MeV as breakdown momentum. Above this scale, the EFT fails to converge.  
\section{\label{data} Data}
The SONIK dataset is a collection of measurements of the $^3$He-$\alpha$ elastic scattering differential cross section for ${}^3$He beam energies between 0.7 and 5.5 MeV performed at the TRIUMF facility, and denoted as experiment SN1687. The experiment is discussed in detail in reference~\cite{paneru_thesis}. The SN1687 experiment used the SONIK apparatus to measure the ${}^3$He-$\alpha$ collision in three interaction regions: interaction region 1 (IR1), interaction region 2 (IR2) and interaction region 3 (IR3). These interaction regions differed slightly in the energy of the collision. The experiment was conducted using nine beam (${}^3$He) energies, with each containing data points corresponding to three slightly different energies in IR1, IR2, and IR3 respectively. In MeV the differential cross section measurements are at  [(0.706, 0.691, 0.676)],[(0.868, 0.854, 0.840)],[(1.292, 1.280, 1.269)],[(1.759, 1.750, 1.741)],[(2.137, 2.129, 2.120)],[(2.624, 2.616, 2.609)], [(3.598, 3.592, 3.586)], [(4.342, 4.337, 4.332)] and [(5.484, 5.480, 5.475)]. Each square bracket is one beam  energy bin with the energies in IR1, IR2 and IR3 from left to right. Note that there were two different runs (of different beam intensities) at the sixth beam energy. Reference~\cite{paneru_thesis} reported measurements in nine energy bins, labeled in keV/u as 239, 291, 432, 586, 711, 873, 1196, 1441 and 1820. In each energy bin we have added an extra $1.8\%$~\cite{private_commune} to the common-mode error provided in reference\cite{paneru_thesis}. We will plot results at these energies, i.e., we do not separate results according to the small energy differences for the different interaction regions. However, when constructing the likelihood the correct energy for data taken in each interaction region was used. 

Apart from the SONIK dataset, we will also be using an analysing power dataset which we will call `Boykin $A_y$' from reference~\cite{boykin}. The reference provides analysing power data for elastic $^3$He-$\alpha$ scattering for three different angles in the centre-of-mass frame, viz; $71.6^\circ$, $87.0^\circ$, and $120.0^\circ$. Results are given at lab energies between $3.30$ MeV and $6.86$ MeV. The data at $71.6^\circ$ and $120.0^\circ$ are actual experimentally measured values while the data at $87.0^\circ$ are extrapolated values from reference~\cite{hardy1968polarization}. The Boykin $A_y$ data however do not include possible systematic uncertainties at all.

There is an older (1964) scattering data by A. C. L. Barnard {\it et al.} that covers the energy range between 2.439-5.741 MeV, and was published in reference~\cite{barnard1964}. We will perform analyses in which this data set is also used to determine ERPs and predict cross sections. When we do that we introduce a (single) additional sampling parameter $f_b$, in order to account for the common-mode error in the Barnard data. The prior\footnote{We will discuss priors in Section \ref{bayesian_formulation}. Meanwhile we write a Gaussian prior on a parameter $\theta$ as $N(\mu_\theta, \sigma_\theta, [a, b])$. Here, the notation simply means that the parameter is distributed normally centered at $\mu_\theta$ with a width of $\sigma_\theta$ but truncated at lower end $a$ and upper end $b$.} on $f_b$ is:
\begin{align}
    \label{prior_on_f_b}
    f_b \sim N(1,\sigma_b,[0.1,2.0]),
\end{align}
where we take the value $\sigma_b=0.05$.
We note that details of the uncertainties in this experiment are not well documented in reference~\cite{barnard1964}.

Another scattering data set (1971) was published by L.~S. Chuang \cite{chuang1971elastic}. It provides differential cross section data at three different lab energies 1.72, 2.46, and 2.98 MeV over an angular range from $34.9^\circ$ to $136.9^\circ$ in the center-of-mass frame. This data set has only 31 points, so on its own it would not yield strong inference on the ERPs. We include it below in concert with other data sets. When we do so we  introduce additional sampling parameters $f_{172}$, $f_{246}$, and $f_{298}$, in order to account for possible common-mode errors in the Chuang data. Since reference~\cite{chuang1971elastic} provides no information on common-mode errors we take the same, essentially uninformative, prior as equation~\eqref{prior_on_f_b} for each of these parameters.

Apart from these data we did not use another  1993 data set of P. Mohr {\it et al.} from reference\cite{mohr1993alpha}. This reference provides differential cross section data at energies in the range $E_{\rm lab}=1.2-3.0$ MeV. Even though the energy range of these data is very favorable for a Halo EFT analysis like ours, they are very inconsistent at the lower energy range, particularly at 1.2 and 1.7 MeV. It is almost impossible to get a reasonable fit (reasonable $\chi^2$) because the very small errors quoted in the data are less than the point-to-point fluctuations in angle. The reference lacks proper discussion of experimental uncertainties and their propagation to the cross sections quoted.

There is also differential cross section data taken by P. D. Miller and G. C. Phillips~\cite{miller1958scattering}. These data cover two different energies at 2.974 and 3.877 MeV. But reference~\cite{miller1958scattering} does not quote any point-to-point and systematic uncertainties. We do not include these data in our analysis.
\section{\label{sec:fwaves}Including the effect of the $\frac{7}{2}^-$ resonance}

\begin{figure}
\centering
\includegraphics[scale=0.5]{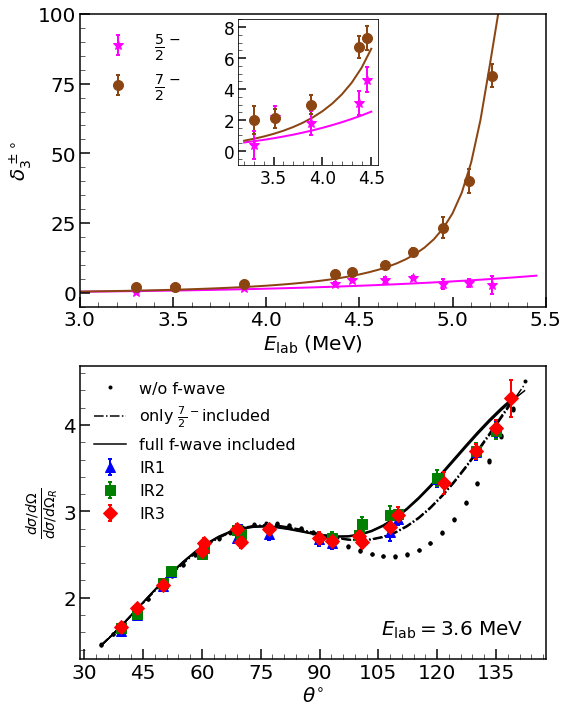}
\caption{Effects of f-waves in the validity region. Top panel: The $\delta^\pm_3$ phase shift from reference~\cite{boykin} demonstrates that the f-wave phase shifts are small but non-zero in the validity region (brown solid dots: $\frac{7}{2}^-$ channel and pink stars: $\frac{5}{2}^-$ channel). The solid lines show the phenomenological {\it R}-matrix calculation using the resonance parameters and channel radius $R_{\rm ch}=4.3$ fm. The zoomed plot shows the upper end of the validity region. Bottom panel: The resonance peak at $5.22$ MeV associated with the ${}^7$Be $\frac{7}{2}^-$ level has a significant effect on the cross section at $E_{\rm lab}=3.6$ MeV, even though this is well below the resonance energy of 5.22 MeV. The dotted line omits both f-wave channels, and the solid line includes them both. The blue triangles, green squares, and red diamonds are SONIK data in each interaction region corresponding to $E_{\rm lab}=3.6$ MeV. The $\frac{5}{2}^-$ resonance has a smaller, but still noticeable effect: omitting the $\frac{5}{2}^-$ partial wave produces the dash-dotted line.}
\label{F-wave_phase}
\end{figure}

The validity region $p\in (60, 90)$ MeV corresponds to lab energy $(2.0, 4.5)$ MeV where the bulk of the SONIK dataset lies. This region avoids the resonance peak associated with the ${}^7$Be $\frac{7}{2}^-$ level at 5.22 MeV. The $\frac{7}{2}^-$ phase shift is known to be small in the validity region, from previous measurements  (see top panel of figure~\ref{F-wave_phase}). Nevertheless, the $\frac{7}{2}^-$ resonance has a noticeable effect on both the cross section and $A_y$ in the validity region. Omitting this channel altogether takes a significant toll on the cross section. This is seen in the bottom panel of figure~\ref{F-wave_phase}, where omitting the f-waves leads to a significant decrease in the cross section at 3.6 MeV (solid line to dotted line).  There is also a smaller, but still noticeable effect in the differential cross section at this energy from the tail of the $\frac{5}{2}^-$ resonance at $9.02$ MeV (width $1.9$ MeV) \cite{tilley2002energy} (solid line vs. dashed line).

The results of figure~\ref{F-wave_phase} imply that for a precision description we need to at least account for the low-energy tail of the $\frac{7}{2}^-$ resonance in the cross section.
In references~\cite{bedaque2003narrow} and~\cite{brown2014field}, narrow resonances near a two-body threshold have been discussed for $\frac{3}{2}^-$ and $\frac{3}{2}^+$ channels respectively in the formalism of EFT. In the future, we look to include the f-wave resonance in our EFT in an analogous way. But, for our purposes here, we found that it is enough to take into consideration the f-wave resonance effects by modeling the two f-wave phase shifts using the phenomenological {\it R}-matrix formula~\cite{brune, laneandthomas}
\begin{eqnarray}
    \label{resonance_eqn}
    \delta_L + \phi_L = \tan^{-1} \Bigg{[}
    -\frac{\Delta_L(E,\rho)}{S_L(E,\rho)}
    \frac{P_L(E,\rho)}{E_L+\Delta_L(E,\rho)-E} 
    \Bigg{]},\\
    \label{1}
    \phi_L = -\tan^{-1} \frac{F_L(\eta,\rho)}{G_L(\eta,\rho)}, \\
    \label{2}
    P_L(E,\rho) = \frac{\rho}{F^2_L + G^2_L}, \\
    \label{3}
    S_L(E,\rho) = P_L \Bigg{[}F_L \frac{dF_L}{d \rho} + G_L \frac{dG_L}{d\rho} \Bigg{]}, \\
    \label{4}
    \Delta_L(E,\rho) = -\frac{\Gamma_L S_L}{2 P_L(E_R,\rho)-\Gamma_L \frac{dS_L(E,\rho)}{dE}|_{E=E_R}},\\
    \label{5}
    E_L=E_R-\Delta_L(E_R,\rho),
\end{eqnarray}
where all quantities are evaluated at $\rho = p R_{\rm ch}$. $R_{\rm ch}$ denotes the channel radius and we choose a fixed channel radius of $4.3$ fm. This choice is enough for our purposes since it produces consistent f-wave phase shifts.

In equation~\eqref{5}, $E_R$ is the resonance peak. $\Gamma_L$ is the resonance width in equation~\eqref{4}.   For the $\frac{7}{2}^-$ level, we fix the resonance peak at 5.22 MeV while taking the width of the peak as a parameter to be determined from sampling. We have used a Gaussian prior for the width:
\begin{align}
    \Gamma_{7/2^-} \sim N(160, 10, [80, 240]).
\end{align} 
We also include the effect of the $\frac{5}{2}^-$ level, for which we fix $E_R=9.02$ MeV and $\Gamma_{\frac{5}{2}^-}=1.9$ MeV. 
\section{\label{bayesian_formulation} Parameter space, Bayesian formulation and MCMC sampling}

The ERPs $a_0, r_0, a^+_{1}, r^+_{1}, P^+_{1}, a^-_{1}, r^-_{1} \ \text{and} \ P^-_{1}$ span an 8-dimensional parameter space. Using equation~\eqref{constraint1}, we
reparameterize the space in terms of the ANCs, replacing $r^\pm_{1}$ by $\mathcal{C}^\pm_1$. We also determine $a^\pm_{1}$ from equation~\eqref{constraint2}, thereby reducing the eight-dimensional parameter space to a six-dimensional. Lastly, we sample $1/a_0$, not $a_0$, since the correlation structure of the parameter space is simpler when it is parameterized in this way. 

Apart from the ERPs, we introduce additional parameters $f_\gamma$ which are normalisation parameters for the differential cross section in the $\gamma$th energy bin of the SONIK dataset. These $f_\gamma$ account for the common-mode error discussed in section 4.4 of reference~\cite{paneru_thesis}.  

Now consider the set of ERPs and normalisation scales together as parameter set $\theta$. Then, we want to construct a credible/confidence interval (CI) from a probability distribution
\begin{equation}
    \label{posterior}    
    p(\{\theta_i\}|D,I)
\end{equation}
which we call the posterior pdf (probability distribution function) for the parameters $\theta_i$ ($i=1,...N_p$). The posterior pdf is simply a function that measures the probability density that the  parameters take a certain value. Such a probability density  can be determined using Bayes' theorem~\cite{bayes}:
\begin{equation}
    \label{bayes}   
    p(\{\theta_i\}|D,I)=\frac{p(D|\{\theta_i\},I) p(\{\theta_i\}|I)}{p(D|I)}.
\end{equation}
The quantity $p(\{\theta_i\}|I)$ is called the prior pdf; it is a distribution function based on our intelligence/ignorance/information \textit{I} on the parameter set $\{\theta_i\}$ prior to any analysis of the given data set \textit{D}. The other quantity $p(D|\{\theta_i\},I)$ is called the likelihood; we denote it by $\mathcal{L}$ below. The likelihood drives the prior pdf towards the credible posterior pdf. $p(D|I)$ is the probability that the given data set \textit{D} is true for the given information \textit{I}; in this work it acts only as a normalisation constant. 

Often in practice we take the logarithm of equation~\eqref{bayes} and so a pragmatic equation is
\begin{equation}
    \label{logposterior}
    \log(p(\{\theta_i\}|D,I)) = \log(p(\{\theta_i\}|I)) + \log(\mathcal{L}) + \text{constant}.
\end{equation}

The results for each parameter, $
\theta_j$ are quoted as a 68\% CI $\bar{\theta}_j\pm \sigma_j$. Here $\bar{\theta}_j$ will be the median of the pdf, and the 68\% interval is determined by numerical integration. For a Gaussian/normal distribution 
this $\bar{\theta}_j$ corresponds to the value of $\theta_j$ that maximizes the posterior pdf (MAP value). The standard deviation $\sigma_j$ for that parameter is then given by
\begin{equation}
    \label{credible_interval}
    \sigma_j^2=-[(\left.\nabla \nabla \log p(\{\theta_i\}|D,I)\right|_{\{\theta_i\}=\{\bar{\theta}_i\}} )^{-1}]_{jj},
\end{equation}
where the quantity denoted by the square brackets $[\cdots]_{jj}$ is the  $j$th diagonal element of the inverse of the Hessian matrix.
 Figures~\ref{parameter_corr_plot} and \ref{norm_corr_plot} show that a Gaussian is an excellent approximation for our final pdf. 

We choose the likelihood pdf to be a gaussian distribution with a correlated chi-squared $\chi^2$, which we write as 
\begin{equation}
    \label{likelihood}
    p(D|\theta,I) \equiv \mathcal{L}=  \frac{1}{\sqrt{(2\pi)^N {\rm det}( \Sigma^\text{expt} + \Sigma^\text{th})}} e^{-\chi^2/2},
\end{equation}
where $N$ is the number of data points and det means the matrix determinant. In equation~\eqref{likelihood}, $\Sigma^\text{expt}_{\alpha \beta}=\sigma^2_\alpha \delta_{\alpha \beta}$
where $\sigma_\alpha$ is the measure of the point-to-point error in the measurement $y_\alpha$ at the $\alpha$th kinematic point. Meanwhile $\Sigma^\text{th}_{\alpha \beta}$ is the theory covariance matrix, defined in our case as \cite{wesolowski2019exploring}
\begin{equation}
    \label{theory_cov}
    \Sigma^\text{th}_{\alpha \beta}=\frac{(y_{\text{ref}})_\alpha (y_{\text{ref}})_\beta c^2_{\text{rms}} Q^{\nu+1}_\alpha Q^{\nu+1}_\beta}{1-Q_\alpha Q_\beta},
\end{equation}
with $(y_{\text {ref}})_\alpha$ the LO Halo EFT cross section at kinematic point $\alpha$,
and $\chi^2$ is defined as
\begin{equation}
    \label{chi2}
    \chi^2 = \vec{r}^T (\Sigma^{\text{expt}}+\Sigma^{\text{th}})^{-1} \vec{r},
\end{equation}
where $\vec{r}$ is the residual matrix. For the $\alpha$th kinematic point the residual is given by
\begin{equation}
\label{residual}
r_\alpha = f_\gamma y_\alpha-Y_\alpha,
\end{equation}
where $y_\alpha$ is the predicted value from theory, $Y_\alpha$ is the measured value at lab and $f_\gamma$ is the aforementioned normalisation factor. (Note that $f_\gamma$ is  fixed for a fixed energy, and so it is not an independent variable for each kinematic point.)

For the ERPs we choose informative priors. In the EFT context we expect the ERPs to have certain sizes. Length-scale ERPs obey naive dimensional analysis with respect to $\Lambda$, although the situation is more complicated for the momentum-scale ERP, $1/a_0$.
We choose Gaussian distribution for these informative priors (numbers are in the appropriate power of ${\rm fm}$ for the parameter):
\begin{align}
1/a_0 & \sim  N(0.02,0.01,[-0.02,0.06]) \\
r_0  & \sim  N(0.0,0.9,[-3.0,3.0]) \\
P^\pm_1 & \sim  N(0.0,1.6,[-6.0,6.0]),
\end{align}

 The priors for ANCs were taken from reference~\cite{zhang2020s} which is an EFT study of capture reaction $^3$He($\alpha$,$\gamma$)$^7$Be that primarily focused on extracting the $S$-factor for the reaction. The reference worked in terms of the sum and ratio of the squared ANCs and its analysis of capture data determined $\mathcal{C}^2_T = {\mathcal{C}^+_1}^2 + {\mathcal{C}^-_1}^2 = 27 \pm 3 \ \text{fm}^{-1}$ and $R = {\mathcal{C}^-_1}^2/{\mathcal{C}^2_T} = 0.48^{+ 0.06}_{-0.05}$. The EFT calculation in reference~\cite{zhang2020s} used the same description of s-wave and p-wave scattering as this study, so
 we build in this information on the ANCs from the reaction 
$^3$He($\alpha$,$\gamma$)$^7$Be by taking~\cite{zhangprivcomm}:
\begin{align}
    {{\cal C}_1^+}^2 &\sim N(13.84,1.63, [5.0,25.0])\nonumber\\
    {{\cal C}_1^-}^2 &\sim 
    N(12.59, 1.85, [5.0,25.0]).
\label{eq:ANCprior}
\end{align}

We also choose gaussian priors for the data normalisation parameters:
\begin{align}
    f_\gamma \sim N(1,\sigma_{f_\gamma},[0.1,2.0])
\end{align}
where the quantities $\sigma_{f_\gamma}$ are taken to be the common-mode error provided for each energy bin in  reference~\cite{paneru_thesis}.

We use \textbf{a}ffine \textbf{i}nvariant \textbf{M}arkov \textbf{C}hain \textbf{M}onte \textbf{C}arlo (aiMCMC), described in detail in references~\cite{foremanemcee, goodmanandweare}, to collect/generate posterior samples distributed according to equation~\eqref{logposterior}. A python implementation of aiMCMC is in the open source package emcee~\cite{foremanemcee}.
\begin{table*}
\caption{\label{bimodal_solution}Bimodal solution of truncated SONIK data: The posterior samples were divided into two regions `lower' where $P^-_1\leq 1.25$ fm and `higher' where $P^-_1>1.25$ fm. Note that the $\chi^2$ values reported are the mimimum $\chi^2$ value in the respective regions; they do not necessarily correspond to MAP values.}
\lineup
\centering
\begin{adjustbox}{width=1\textwidth}
\small
\begin{tabular}{@{}*{11}{l}}
\br
Region & $a_0$ (fm) & $r_0$ (fm) & $a^+_1$ (fm$^3$) & $r^+_1$ (fm$^{-1}$) & $P^+_1$ (fm) & $a^-_1$ (fm$^3$) & $r^-_1$ (fm$^{-1}$) & $P^-_1$ (fm) & $\chi^2$ & $\chi^2/N_{\rm d.o.f.}$ \cr
\mr
Lower & 67 & 0.78 & 243 & \- 0.030 & 1.89 & 163 & \- 0.153 & 0.52 &  310 & 0.81 \cr
Higher & 59 & 0.80 & 163 & \- 0.092 & 1.42 & 335 & \- 0.025 & 1.83 &  317 & 0.83 \cr
\br
\end{tabular}
\end{adjustbox}
\end{table*}
\section{\label{results} Results and Analysis}
\subsection{\label{bimodal_subsection} Bimodality of the posterior samples in the analysis of truncated SONIK data}
After sampling the posterior we found two modes that differed in the p-wave parameters (see figure~\ref{bimodes}). We have tabulated the parameters  corresponding to the MAP values for the two modes in table~\ref{bimodal_solution}. Based on the location of these modes, we divided the posterior samples into two regions; ``lower" and ``higher" based on the value of $P^-_{1}$. We chose $P^-_1$ to make the cut because the posterior samples were  distinctly distributed into two $P_1^-$ regions. The bottom two panels in figure~\ref{bimodes} show that the p-wave phase shifts for the two modes are comparable in magnitude, but result in opposite signs for the p-wave phase-shift splitting. 

\begin{figure*}[!htb]
    \centering
    \includegraphics[height = 5cm, width=\textwidth]{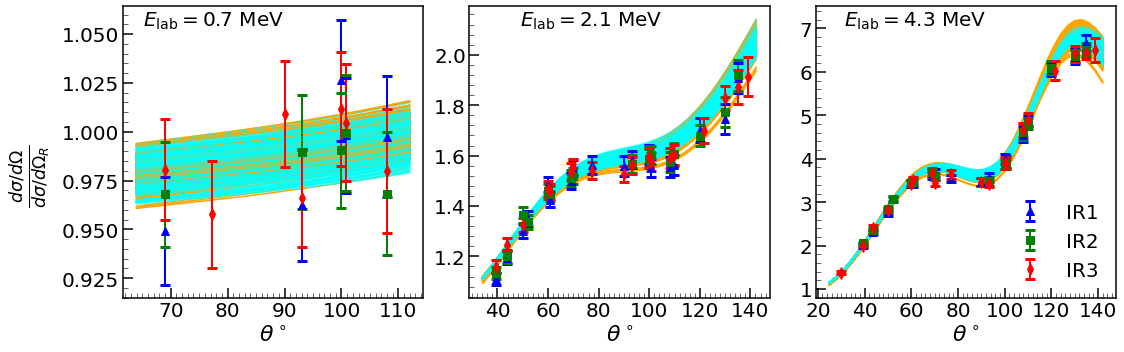}
    \caption{Representative plots of cross sections predicted from the lower (brown) and higher (cyan) region samples (68\% intervals). The plots correspond to three different beam energies, with data obtained in different interaction regions for that beam energy combined. It is clear from the plots that the solutions corresponding to the lower-$P_1^-$ region and the higher-$P_1^-$ region produce essentially identical cross sections.}
    \label{cs_bimodal}
\end{figure*}

The failure of this analysis to determine the sign of the p-wave spin-orbit splitting occurs because that splitting is not directly probed in the cross section. Figure~\ref{cs_bimodal} shows that the lower-$P_1^-$ and higher-$P_1^-$ solutions produce essentially identical cross sections, with both giving a good representation of data. 
They have very similar $\chi^2$ values for the $N=398$ SONIK data points (see last column of table~\ref{bimodal_solution}).  The difference in minimum $\chi^2$s of 7, while not particularly large, does result in only 2\% of our samples lying in the higher-$P_1^-$ region. 
Reference~\cite{barnard1964} also found two p-wave phase-shift solutions that produce comparable cross sections while having different signs for their splitting. Reference \cite{boykin} then obtained a single phase-shift solution when they added their own analysing power data to the phase-shift analysis.

Therefore, in order to resolve this ambiguity in the sign of the p-wave phase-shift splitting, we introduce data on the analysing power,  $A_y(\theta)$ (see equation~\eqref{Ay}) into the likelihood. We add the $\chi^2$ for these data,  
\begin{equation}
    \label{chi2_pol}
    \chi^2_{\text{pol}} = {\vec{r}^T}_{\text{pol}} (\Sigma^\text{expt}_\text{pol})^{-1} {\vec{r}}_{\text{pol}},
\end{equation}
to equation~\eqref{chi2} and collect posterior samples. (Note that theory errors are not included here, as we have not developed a model of order-by-order theory uncertainties for $A_y$.) In equation~\eqref{chi2_pol}
${\vec{r}}_{\text{pol}}$ is the residual matrix for the for $\alpha$th analysing power measurement:
\begin{equation}
\label{residual_pol}
{r_{\text{pol}}}_\alpha = {{y_{\text{pol}}}_\alpha - Y_{\text{pol}}}_\alpha,
\end{equation}
i.e., ${Y_{\text{pol}}}_\alpha$ is the prediction via equation\eqref{Ay} and ${y_{\text{pol}}}_\alpha$ is the experimental measurement of $A_y$, for which we use the Boykin $A_y$ dataset. Since common-mode errors mostly cancel when $A_y$ is computed we do not include a normalisation parameter in equation~\eqref{residual_pol}.  

As expected, upon adding the available analysing power data to the truncated SONIK dataset, we observed that the bimodality is clearly removed and the posteriors settle in the vicinity of higher-$P_1^-$ values. We then obtain the same sign of the spin-orbit splitting as reference~\cite{boykin} does for p-wave phase shifts.

\subsection{Analysis of the truncated SONIK data with truncated analysing power data included}

In this subsection, we present our analysis of a combination of SONIK dataset and the Boykin $A_y$ data. Both data sets are truncated at $E_{\rm lab}=4.3$ MeV. We include the f-wave effects using the phenomenological approach described in section~\ref{sec:fwaves}. In the appendices the correlation plot between all the EFT parameters constrained in our pdf is shown in figure~\ref{parameter_corr_plot} and the posterior pdf of the normalisation parameters is displayed in figure~\ref{norm_corr_plot}. We found that the posteriors for all the parameters are well within the priors that we have set. The MAP values of parameters were already provided in table~\ref{N2LO_trunc_paneru+ay}; this information is repeated in table~\ref{N2LO_all_results}. In this analysis we used 407 data (398 points from SONIK dataset and 9 points from Boykin $A_y$ data) and there were 16 parameters. The minimum $\chi^2$ we obtain corresponds to $\chi^2/N_{\rm d.o.f.}=0.86$. Note that this is the result  for the $\chi^2$ defined in equation~(\ref{chi2}). For the more usual $\chi^2$, computed without theory uncertainties, we get $\chi^2_{\rm expt}/N_{\rm d.o.f.}=0.87$ at the optimum parameter values. The small difference between the two indicates that
\begin{figure*}[!htb]
    \centering
    \includegraphics[height = 10 cm, width=\textwidth]{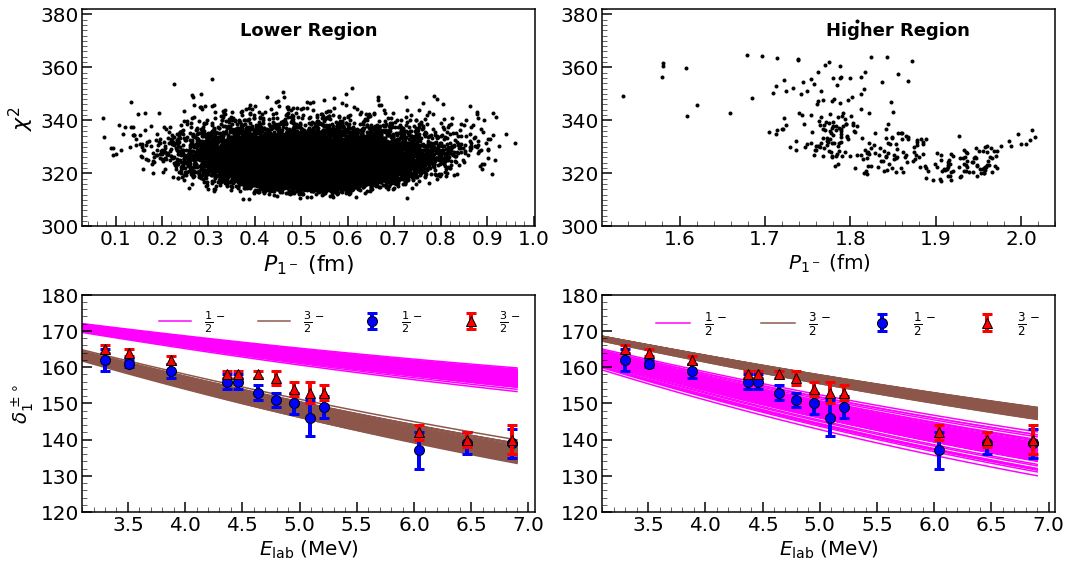}
    \caption{Bimodal solutions to truncated Paneru dataset: The two modes have significant overlap in $\chi^2$ values, as seen in the top panels. These two panels show the distribution of $\chi^2$ as a function of $P^-_1$, with the left (right) panel for the lower (higher) region. The samples were divided into two groups based on a reference value of $P^-_1=1.25$. The bottom two figures are the resulting bands of EFT predicted phase shifts compared to the Boykin phase shifts~\cite{boykin}, which are represented as circles (blue) for the $\frac{1}{2}^-$ channel and triangles (red) for the $\frac{3}{2}^-$ channel. The brown (magenta) band is the EFT prediction for the $\frac{3}{2}^-$ ($\frac{1}{2}^-$) phase shift (68\% credibility interval). 
    }
    \label{bimodes}
\end{figure*}
including the theory uncertainties does not significantly alter the parameter estimation. 

\begin{figure*}
    \centering        \includegraphics[width=\textwidth]{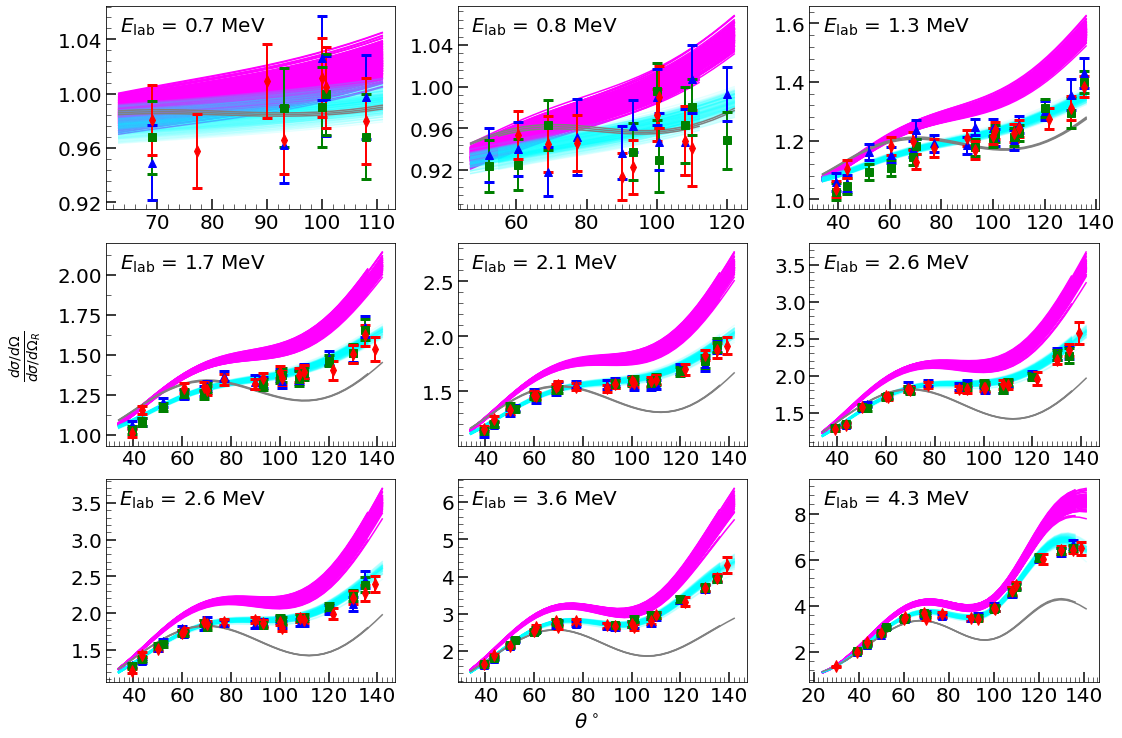} 
    \caption{Predicted band of cross sections at LO, NLO and NNLO: The y-axes in each energy bins are dimensionless cross section, where we have divided by  $\frac{d\sigma}{d\Omega_R}$, the differential cross section for Rutherford scattering. The discrete points are SONIK data in three different interaction regions, viz blue triangles (IR1), green squares (IR2), and red diamonds (IR3). The cross sections are over predicted if we omit the NNLO effects from the calculation and so produce an NLO (magenta band) result from our NNLO samples (cyan band). The cyan band is the full NNLO result. The gray line is the LO result which does not have any parametric uncertainties.}
    \label{cs_at_all_orders}
\end{figure*}

In figure~\ref{cs_at_all_orders} we show bands for the cross section from our posterior samples at each order of the power counting. These 68\% bands only include the parameter uncertainties at each order, they do not show the truncation uncertainty (cf. figure~\ref{fig:validation}). The parameter uncertainty in the cross section is quite small at both NLO and NNLO, while the LO band is just a line since it has no parameter uncertainty. Note that in figure~\ref{cs_at_all_orders} in a single energy bin (one panel), we have constructed separate bands for the three different energies corresponding to IR1, IR2 and IR3. The bands overlap almost completely, so we have plotted all three of those bands using the same color, and labeled each energy bin by a single energy rounded off to one decimal figure. 
Evaluating the NLO cross section with the samples obtained from maximizing the posterior at NNLO (i.e. switching off the NNLO effects) leads to a result that is too large, although the agreement with data is better than at LO. 
However the addition of parameters at NNLO counter-balances the NLO effects and brings the results into good agreement with the data.
\begin{table*}[!t]
\caption{\label{N2LO_all_results}NNLO extraction of parameters for a combination of Barnard dataset({\it B}), SONIK dataset({\it S}) and Chuang dataset({\it C}) with analysing power dataset ($A_y$) truncated at 4.34 MeV. $\chi^2_{\rm red}$ is $\chi^2/N_{\rm d.o.f.}$} 
\lineup
\centering
\begin{adjustbox}{width=1\textwidth}
\small
\begin{tabular}{@{}*{13}{l}}
\br
Data & $a_0$ & $r_0$ & $a^+_1$ & $r^+_1$ & $P^+_1$ & $a^-_1$ & $r^-_1$ & $P^-_1$ & $\Gamma_{{\frac{7}{2}}^-}$ & N & $\text{N}_p$ & $\chi^2_{\rm red}$ \cr
 & (fm) & (fm) & (fm$^3$) & (fm$^{-1}$) & (fm) & (fm$^3$) & (fm$^{-1}$) & (fm) & (keV) &  &  & \cr
\mr
$B$ & 42(6) & 0.85(4) & 174(18) & \-0.072(15) & 1.64(13) & 214(35) & \-0.087(27) & 1.27(30) & 165(8) & 312 & 8 & 0.35 \cr
$SA_y$ & 60(6) & 0.78(2) & 172(5) & \-0.082(4) & 1.50(4) & 288(15) & \-0.044(6) & 1.65(6) & 159(7) & 407 & 16 & 0.86 \cr
$BA_y$ & 42(5) & 0.84(4) & 162(7) & \-0.086(5) & 1.52(5) & 249(16) & \-0.060(6) & 1.56(9) & 166(8) & 321 & 8 & 0.36 \cr
$CA_y$ & \-25(4) & \-0.09(2) & 177(17) & \-0.071(16) & 1.63(13) & 268(25) & \-0.049(13) & 1.65(14) & 159(10) & 40 & 10 & 2.45 \cr
$BSA_y$ & 52(4) & 0.79(2) & 167(5) & \-0.085(3) & 1.50(4) & 276(16) & \-0.047(5) & 1.64(7) & 165(6) & 719 & 17 &  0.67 \cr
$CSA_y$ & 68(7) & 0.77(2) & 172(5) & \-0.081(4) & 1.51(4) & 291(15) & \-0.042(6) & 1.66(6) & 160(7) & 438 & 19 & 1.27 \cr
$BCSA_y$ & 56(4) & 0.78(2) & 167(5) & \-0.085(3) & 1.50(4) & 279(16) & \-0.046(5) & 1.66(7) & 167(6) & 750 & 20 & 0.93 \cr
\br
\end{tabular}
\end{adjustbox}
\end{table*}

\begin{figure}
\centering
\begin{subfigure}{0.5\textwidth}
    \centering
    \includegraphics[height=0.6\textheight, width=1.0\textwidth]{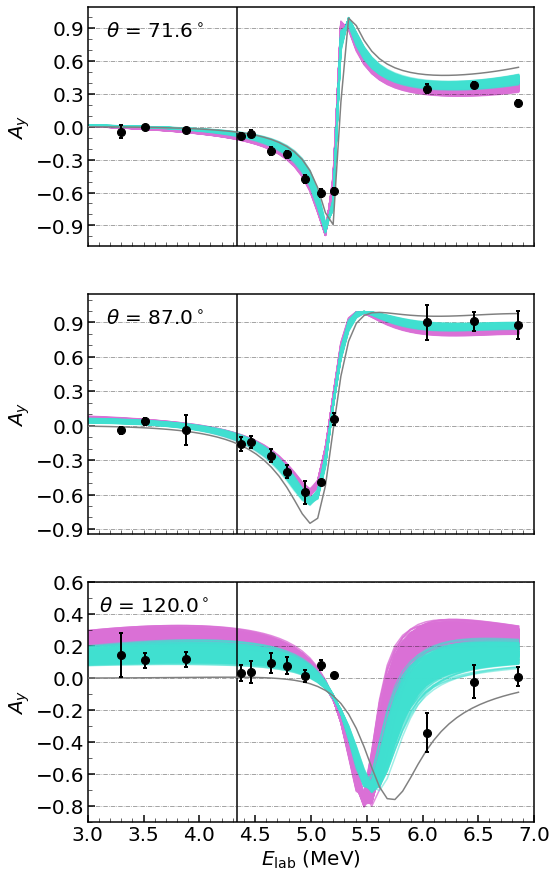}
    \subcaption{Predicted analysing power}
    \label{fig:analysing_power}
\end{subfigure}%
\begin{subfigure}{0.5\textwidth}
    \centering
    \includegraphics[height=0.6\textheight,width=1.0\textwidth]{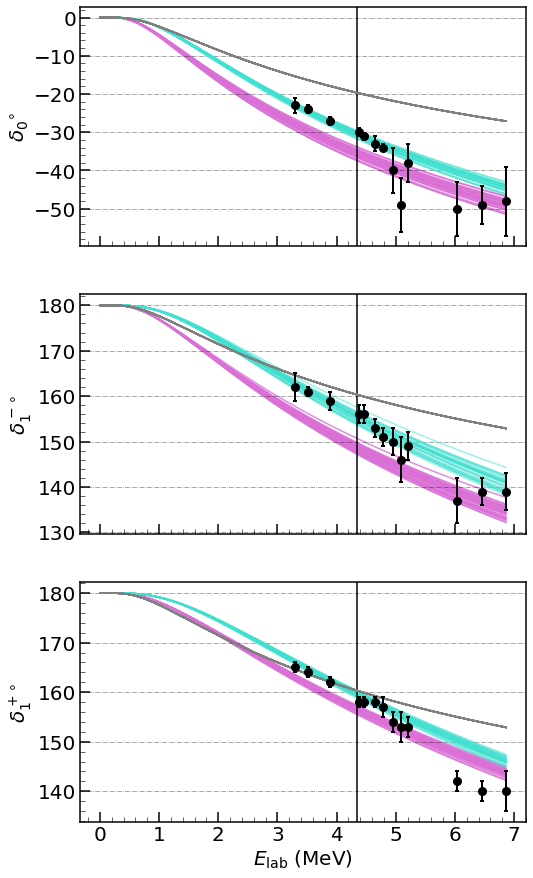}
    \subcaption{Predicted phase shifts}
    \label{fig:phase_pred_N2LO}
\end{subfigure}
\caption{Predicted analysing power and phase shifts at different orders: The dark vertical line is the truncation point ($E_{\rm lab}=4.3$ MeV) of the SONIK dataset. The band (LO=gray, NLO=pink, NNLO=cyan) represents the predicted values of at different orders obtained from the NNLO  analysis of truncated $SA_y$ dataset.}
\end{figure}

We also plot the EFT result for the analysing power and compare it with the available Boykin $A_y$ data (see figure~\ref{fig:analysing_power}). Note that here we used only nine of the available 39 data points (the first three at each available angle). It is already impressive that even only a few analysing power data are sufficient to fix the sign of phase-shift splitting (see subsection \ref{bimodal_subsection} and figure~\ref{bimodes}).

We used the NNLO analysis to extract the phase shifts for s- and p-waves (see figure~\ref{fig:phase_pred_N2LO}). Up to the maximum energy ($E_{\rm lab}=4.3$ MeV) considered in this analysis (vertical black line) our extracted phase shifts for s- and p-waves are consistent with the 
Boykin phase shifts within the uncertainties. Especially for s-waves, above $E_{\rm lab}=4.3$ MeV the Boykin phase shifts contain sizable fluctuations and have sizable error bars. We suggest only a qualitative comparison with our phase shifts there, and at this level the agreement is good.

We found that the parameters $a_0$ and $r_0$ are anti-correlated.  $a_0$ comes out somewhat larger than our scale assignment $\Lambda^2/p_{\rm typ}^3$ and $r_0$ come out  smaller than the NDA estimate $1/\Lambda$. Nevertheless, the $a_0$-$r_0$ posterior is well within the prior chosen. Our central value for $a_0$ ($r_0$) is larger (smaller) than previous values (cf. subsection~\ref{sec:previouswork}).     

Apart from the s-wave ERPs, the way we have performed our sampling facilitated the calculation of p-wave ANCs and the resonance width of the $\frac{7}{2}^-$ level. We found: 
\begin{align} 
{\mathcal{C}^+_{1}}^2 &= 15.5 \pm 1.5 \ \text{fm}^{-1} \\ 
{\mathcal{C}^-_{1}}^2 &= 14.1 \pm 1.7 \ \text{fm}^{-1}.
\label{eq:ANCresult}
\end{align}
Our analysis only included data up to 4.3 MeV and so could not tightly constrain the resonance width of $\frac{7}{2}^-$ level. We found $\Gamma_{\frac{7}{2}^-} = 159\pm7$ keV, which is only a bit narrower than the prior, but emphasizes the lack of adequate resonance information in the truncated data. Because of this limitation in the truncated data, the phenomenological inclusion of the resonance is sufficient for our purposes.
 
Lastly, we find that the normalisation factors for each SONIK energy bin were well inside the common-mode errors quoted in reference~\cite{paneru_thesis} although the normalisation factors for energy bins above ~2.2 MeV push towards the upper limit. This is depicted in figure~\ref{norm_corr_plot}, shown in the supplemental material.

\subsection{Adding Barnard Data to analysis}
We next include the data of reference~\cite{barnard1964} which we call Barnard data ({\it B}) in our analysis, truncating this data set also at 4.342 MeV. In doing so we introduced an additional sampling parameter $f_b$ which accounts for the common-mode error in the Barnard data.

The values of ERPs extracted from this analysis are tabulated in table~\ref{N2LO_all_results}. Surprisingly, we find that the posterior pdf when sampling the truncated Barnard data is unimodal even though both Barnard {\it et al}'s original paper~\cite{barnard1964} and the phase-shift analysis by Boykin~\cite{boykin} found two modes that correspond  to the two modes that we found when sampling the truncated SONIK data. 

Analysing the Barnard data alone yields a smaller scattering length ($a_0$), larger effective range ($r_0$) and broader uncertainties in other parameters compared to an analysis of the SONIK data. The Halo EFT fit has a smaller $\chi^2$ to the Barnard data compared to the fit to the SONIK data. Indeed the $\chi^2/N_{\rm d.o.f.}$ once the Barnard data is included is essentially too small, suggesting that the larger point-to-point uncertainties reported for each of the data points in the Barnard set may be too large. 

\subsection{Adding Chuang Data to analysis}

We also analysed the Chuang data ({\it C}) of reference~\cite{chuang1971elastic} together with the truncated (at $E_{\rm lab} = 4.3$ MeV) $A_y$ data and SONIK data. In contrast to other analyses, the Chuang and $A_y$ data, yield negative MAP values for parameters $a_0$ and $r_0$ (see table \ref{N2LO_all_results}). The uncertainties in $a_0$ and $r_0$ from this analysis are of similar size to other cases, but the uncertainties in other parameters are much larger. The $\chi^2/N_{\rm d.o.f}$ is also markedly bigger than the $\chi^2/N_{\rm d.o.f}$ in other analyses. The small number of Chuang data and the relatively small energy range mean that the ERPs come out much more correlated with one another in the $CA_y$ analysis than once the SONIK data is included. The results of the $CSA_y$ analysis are dominated by the SONIK dataset, and yield ERPs that are very similar to the $SA_y$ analysis.

\subsection{\label{sec:previouswork}Comparison to previous Halo EFT analyses}
In reference~\cite{higa2018radiative} Higa {\it et al} analysed data on the radiative capture reaction ${}^3{\rm He}(\alpha,\gamma){}^7{\rm Be}$ as well as the phase shifts of Boykin et al.~\cite{boykin}. The NLO result of Higa {\it et al} has the same parameters as our NNLO analysis: $a_0$ and $r_0$ in the s-wave and terms through order $p^4$ in the effective-range expansion of both p-wave amplitudes. But, in addition, they include a shape parameter in the s-wave. Higa {\it et al} find that the Boykin phase shifts give $P_1^+=1.59 \pm 0.03$ fm and $P_1^-=1.74 \pm 0.05$ fm. The former overlaps with our result at 1.5$\sigma$ and the latter is consistent with ours within the 1$\sigma$ error bar. The s-wave scattering length $a_0$ found in this analysis is much smaller than ours, at $21.6 \pm 3.4$ fm, and the effective range is larger: reference~\cite{higa2018radiative} quotes $r_0=1.2 \pm 0.1$ fm. A subsequent Bayesian analysis by Premarathna and Rupak gives revised numbers, including, for the s-wave parameters, $a_0=40^{+5}_{-6}$ fm and $r_0=1.09^{+0.09}_{-0.1}$ fm~\cite{premarathna2020bayesian}. We reiterate that these s-wave ERPs are extracted from Boykin phase shifts that are obtained from data which only extends down to $E_{\rm lab}=3.3$ MeV. In contrast, our numbers for $a_0$ and $r_0$ are based on directly analysing SONIK data that extends down to $E_{\rm lab}=676$ keV. 

The value of $a_0=60 \pm 6$ fm found here from the SONIK data is consistent with the analysis of ${}^3$He($\alpha$,$\gamma$) data in reference~\cite{zhang2020s}, which concluded $a_0=50^{+7}_{-6}$ fm. We note that the analysis conducted in our work includes only ${}^3$He-$\alpha$ scattering data: it does not include any of the capture data considered in reference~\cite{zhang2020s}. It is gratifying that the results for $a_0$ are consistent. In contrast, the value obtained for $r_0$ in reference~\cite{zhang2020s} is $0.97 \pm 0.03$ fm, markedly larger than the $0.78 \pm 0.02$ fm found here using the scattering data.

In fact, we do effectively include some information on ${}^3$He($\alpha$,$\gamma$), since we take the two-dimensional posterior on the p-wave ANCs found in reference~\cite{zhang2020s} and use it as a prior for this analysis. The resulting p-wave ANCs are almost as tightly constrained at the 1$\sigma$ level as in the prior, but the central values are shifted slightly larger. The ANC prior and posterior do overlap at the 1$\sigma$ level, cf. equations~(\ref{eq:ANCresult}) and (\ref{eq:ANCprior}), so there is no conflict between the SONIK data and the ANCs needed to fit the capture data.
\section{\label{conclusion} Conclusion and Summary}
The need for accurate low-energy data for $^3$He-$\alpha$ scattering in order to better understand the ${}^3$He($\alpha$,$\gamma$) reaction at solar energies has existed for a long time. In order to meet this need an experiment on elastic scattering of $^3$He-$\alpha$ has recently been carried out using SONIK gas target at TRIUMF, reaching center-of-mass energies as low as 0.3 MeV. This scattering has been analysed using {\it R}-matrix by Paneru and the data (referred to here as the SONIK data) are presented in reference~\cite{paneru_thesis} and the forthcoming reference~\cite{Paneru:2021}. 

In this paper we have constructed an EFT for this scattering process that is consistent in the momentum (lab) validity region between $60-90$ MeV where the bulk of the SONIK data lies. We have demonstrated that in this region we can expand the differential cross section in powers of the expansion parameter, $Q$, that corresponds to the ratio of the typical momentum ($p_{\text{typ}}$) of the collision and a breakdown scale $\Lambda=200$ MeV. We organized the ERPs that characterize the s-wave and p-wave amplitudes. All the momentum-scale ERPs  ($\frac{1}{a_0}, \frac{1}{a^\pm_1}, r^\pm_1$) are unnaturally small compared to what is expected on the basis of naive dimensional analysis with respect to $\Lambda$, while all the length-scale ERPs ($r_0, P^\pm_1$) tend to be natural when expressed in units of $1/\Lambda$. Based on this observations, we proposed a power counting scheme in which the LO contributions come solely from the Coulomb-modified short range interactions and s-and p-wave ERPs play roles at NLO and NNLO. We demonstrated this is a consistent organization of the problem in the validity region. In this hierarchy of mechanisms  the NLO result over-predicts the data, but it is corrected through the proper inclusion of additional parameters at NNLO. We examined other power-counting schemes that could be deemed suitable; however all lead to similar overall results at NNLO. Ultimately, we choose the particular power counting tabulated in table \ref{power_counting_table} because this hierarchy of mechanisms leads to EFT expansion coefficients, $c_i$, of the differential cross section that are of similar size at LO, NLO, and NNLO. This allows us to formulate a statistical model for the N$^3$LO coefficient and generate uncertainties that account for the error induced by truncating the EFT expansion at NNLO. 

By formulating a likelihood that includes this theory uncertainty and using MCMC sampling, we collected posterior samples of ERPs at NNLO.  Table \ref{N2LO_all_results} provides the central values and uncertainties of the ERPs corresponding to our Bayesian posterior. The result we find for the s-wave scattering length from the SONIK data, $a_0=60 \pm 6$ fm, is consistent with that obtained from analysing ${}^3{\rm He}(\alpha, \gamma){}^7{\rm Be}$-capture data in the same EFT~\cite{zhang2020s}, although the values of $r_0$ are quite different. The disagreement between different analyses regarding the value of $r_0$ indicates that experimental data which clarifies the energy dependence of the s-wave phase shift would be very valuable. Including the Barnard data~\cite{barnard1964} in the analysis yields a slightly smaller central value for $a_0$, but the results are consistent within error bars. It also leads to somewhat different ANCs (see figure\ref{fig:compare_results}). In contrast, analysing the data of reference~\cite{chuang1971elastic} together with the SONIK data yields a slightly higher---but still consistent---$a_0$. The value $a_0=60 \pm 6$ fm we obtain from our analysis of SONIK data is not in conflict with the data from the two older experiments we examined here.

In some previous works, optical potentials have been used to model this reaction. For example, in reference~\cite{mohr1993alpha} Mohr {\it et al.} fit optical model parameters to the ${}^3{\rm He}(\alpha,\gamma){}^7{\rm Be}$-capture data along with elastic phase shifts. Direct comparison of Mohr {\it et al}'s phase shifts with the ones we extracted above shows they agree within the errors. This immediately reveals the fact that our phase shifts are consistent with the spin-orbit splitting in Mohr \textit{et al}'s optical potential. That optical potential also produces the correct splitting between the ${\frac{1}{2}}^-$ and ${\frac{3}{2}}^-$ bound states. Furthermore, since the Mohr {\it et al}. phase shifts and our phase shifts agree within uncertainties the Mohr {\it et al} optical potential should produce similar ERPs to the ones obtained in our study.

Since we have a working and consistent theoretical error model included in this analysis, we can use it to extrapolate the amplitude $t^{\rm CS} = \sum_L T_L(E)$ towards lower energies. However, one has to be very careful in extending such extrapolation to the cross section. As the collision energy goes to zero higher-order electromagnetic effects like relativistic corrections to the Coulomb potential, magnetic-moment interactions, vacuum polarization, polarizability effects etc. will play a role. Including these effects will facilitate consistent and precise extrapolation, but is beyond the scope of this work.  

The inclusion of analysing-power data in the likelihood is a crucial step in obtaining these results: it is needed to  resolve the sign of the p-wave phase-shift splitting at lower energies in this system. Even though the p-wave contributions to cross sections get smaller as the scattering energy decreases, analysing-power data plays a vital role in deducing the sign ambiguity. It would be very useful to have more polarisation data at lower energies: this would aid future analyses like ours.

In particular, we have included the f-wave resonance in our analysis because of its significant effect on the cross section at the uppermost energies in our analysis. But one may choose to omit the f-wave influence at energies less than $E_{\rm lab}=3.0$ MeV if more data at such energies would be available in the future. Such lower-energy data will also help to constrain the s-wave ERPs better.  

This inclusion of the f-wave resonance was done in a purely phenomenological way here, using the {\it R}-matrix formalism of Lane and Thomas~\cite{laneandthomas}. Future work to formulate an EFT of f-wave resonances could remove this inconsistency and should allow the EFT to be used to analyse scattering data all the way from threshold through the $\frac{7}{2}^-$ resonance. 

The use of both scattering and capture data in an EFT analysis of the ${}^7$Be system is also an important topic for future work. While some information from the capture data was included here through the use of a prior on the ${}^7$Be ANCs, this is only part of the full posterior pdf. 

Lastly, we mention that the extension of the EFT worked out here to N$^3$LO is straightforward: at that order three new ERPs enter, all of dimension fm$^3$, in the $\frac{1}{2}^+$, $\frac{1}{2}^-$, and $\frac{3}{2}^-$ channels. Provided these shape parameters are natural with respect to $\Lambda$ their impact on the cross section should be encompassed within the theory error bars shown in figure~\ref{fig:validation}. 
\ack 
We thank Carl Brune, Daniel Odell, Somnath Paneru, and James deBoer for many valuable discussions and suggestions,  Charlotte Elster for initially alerting us to the importance of polarisation data, and Chlo\"e Hebborn for useful discussions on the Coulomb Green's function and Gautam Rupak for clarifying information regarding the analyses of references~\cite{higa2018radiative, premarathna2020bayesian}. This work has been funded by US Department of Energy, Office of Science, Office of Nuclear Physics under Awards DE-FG02-93ER-40756, by the National Nuclear Security Agency under Award DE-NA0003883, and by the US NSF via grant PHY-2020275, N3AS PFC.
\clearpage
\appendix
\counterwithin{figure}{section}
\section{Full parameter correlation plot}
\begin{minipage}{\textwidth}
    \centering
    \label{parameter_corr_plot}
    \includegraphics[height=0.76\vsize, width=1.0\hsize]{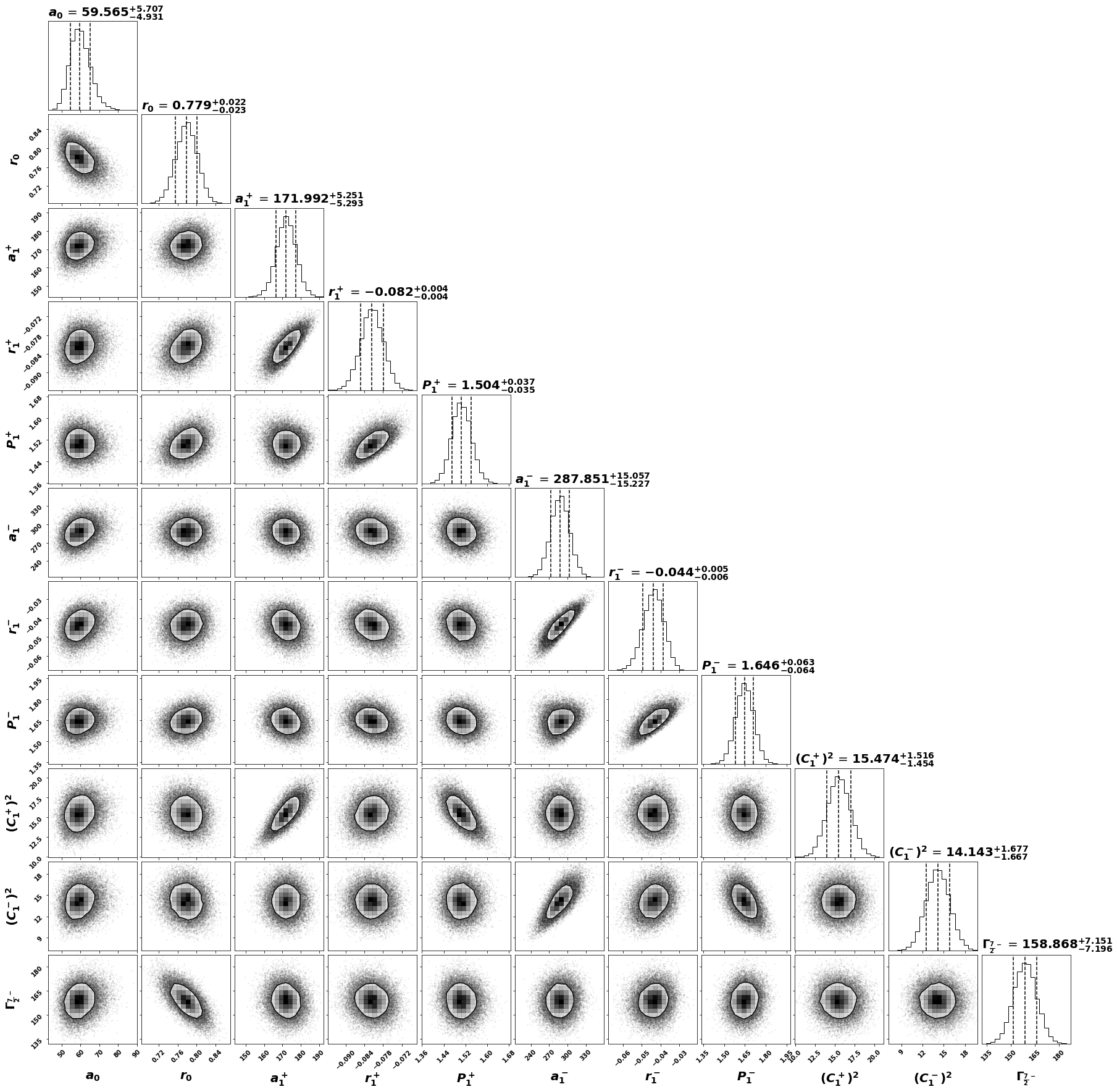}
    \captionof{figure}{Correlation plot of EFT and resonance parameters from sampling $SA_y$ data, truncated at $E_=4.3$ MeV (lab). The values quoted are the median values of the posterior samples.}
\end{minipage}

\clearpage

\section{Normalisation correlation plot}
\noindent
\begin{minipage}{\textwidth}
    \centering
    \label{norm_corr_plot}
    \includegraphics[height=0.76\vsize, width=1.0\hsize]{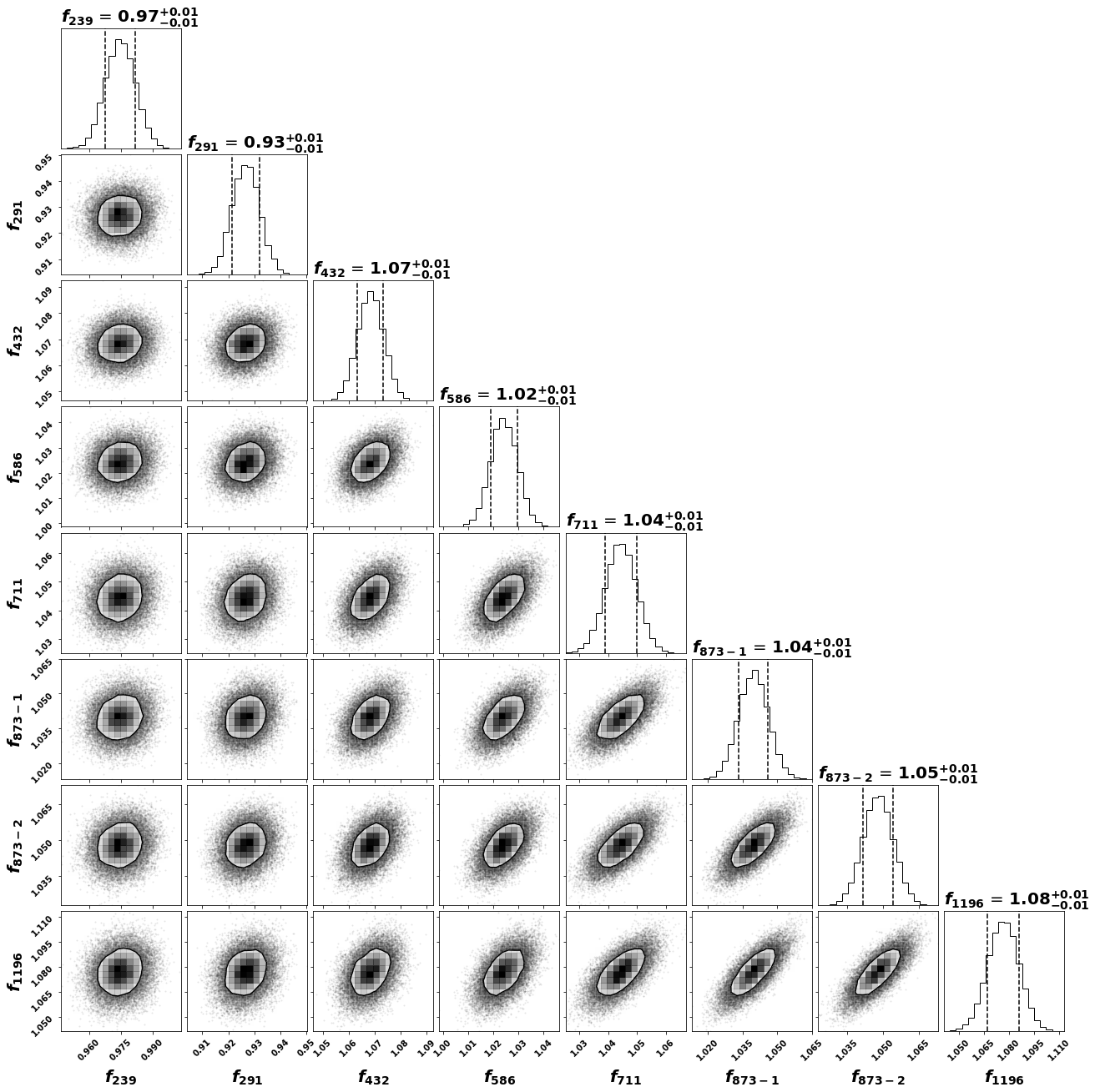}
    \captionof{figure}{Correlation plot of normalisation parameters from sampling $SA_y$ data truncated at $E=4.3$ MeV (lab). The values quoted are the median values of the posterior samples.}
\end{minipage}

\clearpage

\section{Comparison of central values of the ERPs}
\noindent
\begin{minipage}{\textwidth}
    \centering
    \includegraphics[width=15.0 cm]{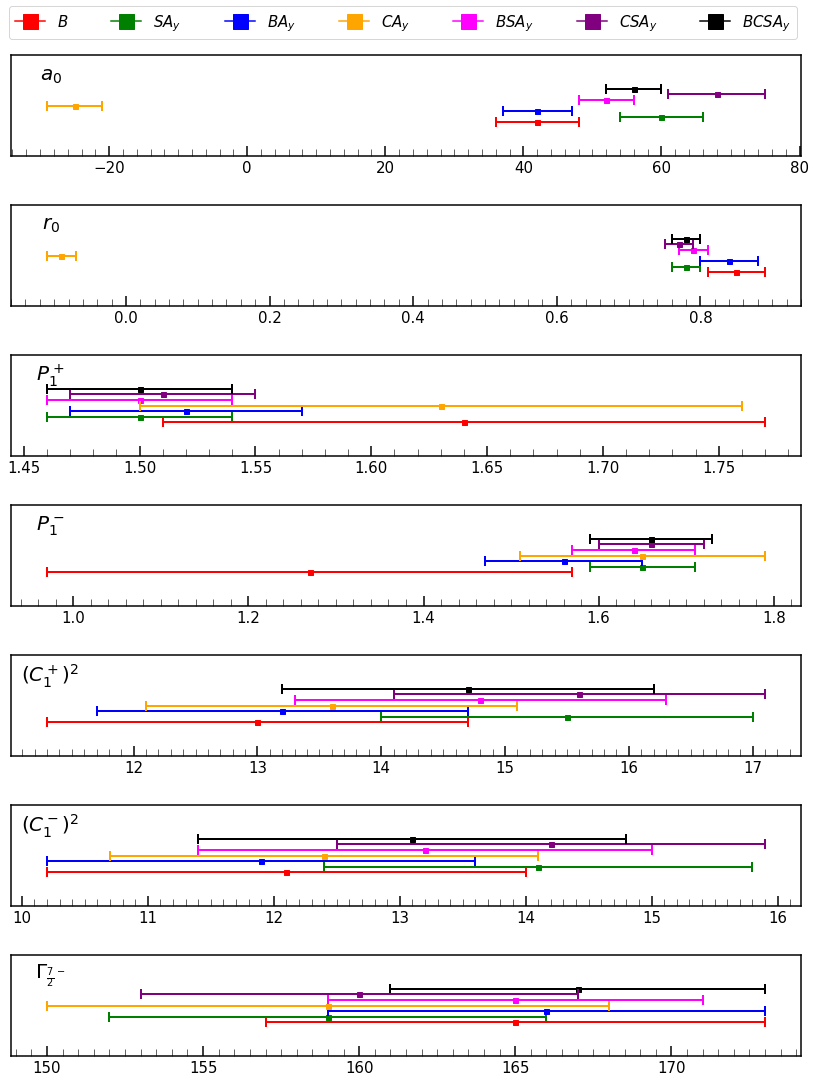}
    \captionof{figure}{Comparison of the central values of the parameters extracted from analyses of different data sets. This is also a pictorial representation of some of the results listed in table~\ref{N2LO_all_results} with ANCs added here.}
    \label{fig:compare_results}
\end{minipage}

\clearpage

\section{Central values of the normalisation parameters}
\begin{minipage}{\textwidth}
\captionof{table}{Normalisation parameters obtained from sampling SONIK data with other datatset} 
\lineup
\centering
\begin{adjustbox}{width=1\textwidth}
\small
\begin{tabular}{@{}*{10}{l}}
\br
Data & $f_{239}$ & $f_{291}$ & $f_{432}$ & $f_{586}$ & $f_{711}$ & $f_{873-1}$ & $f_{873-2}$ & $f_{1196}$ & $f_{1441}$ \cr
\mr
$S$ (lower mode) & 0.97(0.01) & 0.93(0.01) & 1.07(0.01) & 1.02(0.01) & 1.03(0.01) & 1.03(0.01) & 1.03(0.01) & 1.05(0.01) & 1.08(0.01) \cr
$S$ (higher mode) & 0.97(0.01) & 0.93(0.01) & 1.07(0.01) & 1.02(0.01) & 1.04(0.01) & 1.03(0.01) & 1.03(0.01) & 1.06(0.01) & 1.08(0.01) \cr
$SA_y$ & 0.97(0.01) & 0.93(0.01) & 1.07(0.01) & 1.02(0.01) & 1.04(0.01) & 1.04(0.01) & 1.04(0.01) & 1.05(0.01) & 1.08(0.01) \cr
$BSA_y$ & 0.97(0.01) & 0.93(0.01) & 1.07(0.01) & 1.02(0.01) & 1.04(0.01) & 1.03(0.01) & 1.03(0.01) & 1.06(0.01) & 1.08(0.01) \cr
$CSA_y$ & 0.98(0.01) & 0.93(0.01) & 1.07(0.01) & 1.03(0.01) & 1.05(0.01) & 1.05(0.01) & 1.05(0.01) & 1.09(0.01) & 1.11(0.01) \cr
$BCSA_y$ & 0.98(0.01) & 0.93(0.01) & 1.07(0.01) & 1.03(0.01) & 1.05(0.01) & 1.04(0.01) & 1.05(0.01) & 1.08(0.01) & 1.10(0.01) \cr
\br
\end{tabular}
\end{adjustbox}

\caption{Normalisation parameters obtained from sampling Barnard data with other datatset} 
\lineup
\centering
\small
\begin{tabular}{@{}*{2}{l}}
\br
Data & $f_b$ \cr
\mr
$B$ & 1.05(0.01) \cr
$BA_y$ & 1.05(0.01) \cr
$BSA_y$ & 1.05(0.01) \cr
$BCSA_y$ & 1.05(0.01) \cr
\br
\end{tabular}

\caption{Normalisation parameters obtained from sampling Chuang data with other datatset} 
\lineup
\centering
\small
\begin{tabular}{@{}*{4}{l}}
\br
Data & $f_{172}$ & $f_{246}$ & $f_{298}$ \cr
\mr
$CA_y$ & 1.04(0.01) & 1.03(0.03) & 0.99(0.03) \cr
$CSA_y$ & 1.04(0.01) & 0.97(0.02) & 0.82(0.02) \cr
$BCSA_y$ & 1.04(0.01) & 0.97(0.02) & 0.81(0.02) \cr
\br
\end{tabular}

\end{minipage}

\clearpage
\section*{References}
\bibliographystyle{unsrt}
\bibliography{references}
\end{document}